%% file: main.tex
\documentclass[conference]{IEEEtran}
\IEEEoverridecommandlockouts

\usepackage{amsmath,amsfonts} 
\usepackage{algorithmic}
\usepackage{graphicx}
\usepackage{textcomp}
\usepackage{xcolor}
\usepackage{enumitem,kantlipsum}
\usepackage{subcaption}
\usepackage[hidelinks]{hyperref}
\usepackage{cleveref}
\usepackage{algorithm}
\usepackage{microtype}
\usepackage{cancel}
\usepackage{xspace}
\usepackage{xfrac}
\usepackage{units}
\usepackage{verbatim}

\newcommand{\TRNA}{{$\text{\emph{TuNA}}$}\xspace}
\newcommand{\HTRNA}[2]{{$\text{\emph{TuNA}}_\text{\emph{\scriptsize #1}}^\text{\emph{\scriptsize #2}}$}\xspace}

\newcommand{\HTRNATS}{{\HTRNA{l}{g}}\xspace}

\usepackage{etoolbox}
\newtoggle{reviewMode}

\def\BibTeX{{\rm B\kern-.05em{\sc i\kern-.025em b}\kern-.08em
    T\kern-.1667em\lower.7ex\hbox{E}\kern-.125emX}}
\begin{document}


\title{Configurable Non-uniform All-to-all Algorithms}

\author{
    \IEEEauthorblockN{Ke Fan\IEEEauthorrefmark{1}, Jens Domke\IEEEauthorrefmark{2}, Seydou  Ba\IEEEauthorrefmark{2}, and Sidharth Kumar\IEEEauthorrefmark{1}}
    \IEEEauthorblockA{\IEEEauthorrefmark{1}University of Illinois Chicago, Chicago, IL, USA\\
    Email: \{kfan23, sidharth\}@uic.edu}
    \IEEEauthorblockA{\IEEEauthorrefmark{2}RIKEN Center for Computational Science, Kobe, Japen\\
    Email: \{jens.domke, seydou.ba\}@riken.jp}
}
\maketitle

\begin{abstract}
\texttt{MPI\_Alltoallv} generalizes the uniform all-to-all communication (\texttt{MPI\_Alltoall}) by enabling the exchange of data blocks of varied sizes among processes. This function plays a crucial role in many applications, such as FFT computation and relational algebra operations.
Popular MPI libraries, such as MPICH and OpenMPI, implement \texttt{MPI\_Alltoall} using a combination of linear and logarithmic algorithms. 
However, \texttt{MPI\_Alltoallv} typically relies only on variations of linear algorithms, missing the benefits of logarithmic approaches. Furthermore, current algorithms also overlook the intricacies of modern HPC system architectures, such as the significant performance gap between intra-node (local) and inter-node (global) communication. 
This paper introduces a set of \emph{Tu}nable \emph{N}on-uniform \emph{A}ll-to-all algorithms, denoted \HTRNA{l}{g}, where \emph{g} and \emph{l} refer to global (inter-node) and local (intra-node) communication hierarchies.
These algorithms consider key factors such as the hierarchical architecture of HPC systems, network congestion, the number of data exchange rounds, and the communication burst size. 
The algorithm efficiently addresses the trade-off between bandwidth maximization and latency minimization that existing implementations struggle to optimize. 
We show a performance improvement over the state-of-the-art implementations by factors of $42$x and $138$x on Polaris and Fugaku, respectively.
\end{abstract}

\input{introduction}
\input{trav}

\input{ltrav}

\input{evaluation}
\input{application}

\input{related_work}

\input{conclusion}

\bibliographystyle{IEEEtran}
\bibliography{main}

\end{document}

%% file: introduction.tex
\vspace{1em}
\section{Introduction}\label{sec:intro}

Optimizing data movement remains a critical challenge in the era of exascale.
Collective communication that involves data exchange among (almost) all processes is an important class of data movement. Owing to its global scope, collectives are typically difficult to scale, and can consume a substantial portion of the overall execution time in applications, often accounting for between $25\%$ and $50\%$, or more~\cite{bernholdt2020survey}. 
Machine learning (ML) applications, in particular, depend heavily on all-reduce and all-to-all (both \emph{uniform} and \emph{non-uniform}) collectives, which are crucial in efficiently shuffling data and synchronizing parameters during the parallel training process~\cite{chen_highly_2022,zhou_accelerating_2024}.
Beyond ML, various HPC workloads heavily utilize \emph{non-uniform} all-to-all communication. These include graph algorithms like PageRank~\cite{besta_push_2017}, Fast Fourier Transform (FFT) computations~\cite{nvidia_corporation_multinode_2022}, quantum computer simulations~\cite{willsch_large-scale_2023}, and certain advanced preconditioners and solvers~\cite{collom_optimizing_2023}.


State-of-the-art implementations of \emph{non-uniform} all-to-all communication typically rely on linear-time algorithms. In contrast, \emph{uniform} all-to-all collective implementations utilize either linear-time algorithms (e.g., the scattered algorithm~\cite{mpich-web}) or logarithmic-time algorithms (e.g., the Bruck algorithm~\cite{thakur2005optimization}), depending on the message sizes.
Adapting logarithmic-time algorithms for \emph{non-uniform} all-to-all communication is a challenging task and has only recently been explored~\cite{fan2022optimizing}.
However, while being logarithmic, this work only works for a fixed base of 2 and is agnostic of the architecture hierarchy of HPC systems, leaving substantial room for further optimization. 



Our research focuses on improving the efficiency of all-to-all communication for \emph{non-uniform} data distributions, addressing the limitations of existing approaches. We introduce a novel algorithm called \TRNA (\textbf{tu}nable-radix \textbf{n}on-uniform \textbf{a}ll-to-all), specifically designed for \emph{non-uniform all-to-all} workloads. \TRNA's key innovation lies in its adjustable radix, which determines the base of the logarithmic complexity. This radix can be set anywhere between 2 and the total number of processes ($P$). By allowing fine-grained control over the radix, \TRNA enables users to optimize the trade-off between the number of communication rounds and the size of data exchanges, leading to improved performance scalability.


Furthermore, the recent increase in CPU cores and NUMA domains requires some HPC workloads to be executed in an $m \times n$ configuration per compute node (w.r.t MPI ranks and OpenMP threads)~\cite{domke_a64fx_2021} to utilize the available resources effectively.
To address the communication needs of these workloads, we have extended \TRNA with hierarchical variants, referred to as \HTRNA{l}{g}. This extension separates the all-to-all communication into two distinct components: intra-node (local) and inter-node (global). By doing so, \HTRNA{l}{g} can better adapt to the multi-level structure of contemporary HPC systems, potentially improving overall communication efficiency.


In summary, our paper makes the following contributions:

\begin{enumerate}[leftmargin=*]
    \item We develop \TRNA, capable of adjusting the radix ($r$) from~$2$ to~$P$. Evaluation of \TRNA reveals: small radices work for small messages, a radix close to~$\sqrt{P}$ improves mid-sized communication, and large radices work for large messages.
    \item We develop hierarchical \TRNA and its variants that decouple the communication into local intra-node and global inter-node data exchange phases, to improve performance further.
    \item We perform a detailed evaluation of our techniques using scaling studies (up to $16$k processes) on Fugaku~\cite{sato_co-design_2020} and Polaris~\cite{argonne_national_laboratory_polaris_2024}. 
    Our algorithms exhibit a performance improvement of $60.60\times$ (\TRNA), $138.59\times$ (\HTRNA{l}{g}) over the vendor implementation of \texttt{MPI\_Alltoallv}.
\end{enumerate}

\section{Background}\label{sec:background}

\begin{figure}[tbp]
    \centering
    \vspace{-1.2em}
    \includegraphics[width=1\linewidth, trim = 5cm 6cm 6cm 5cm, clip]{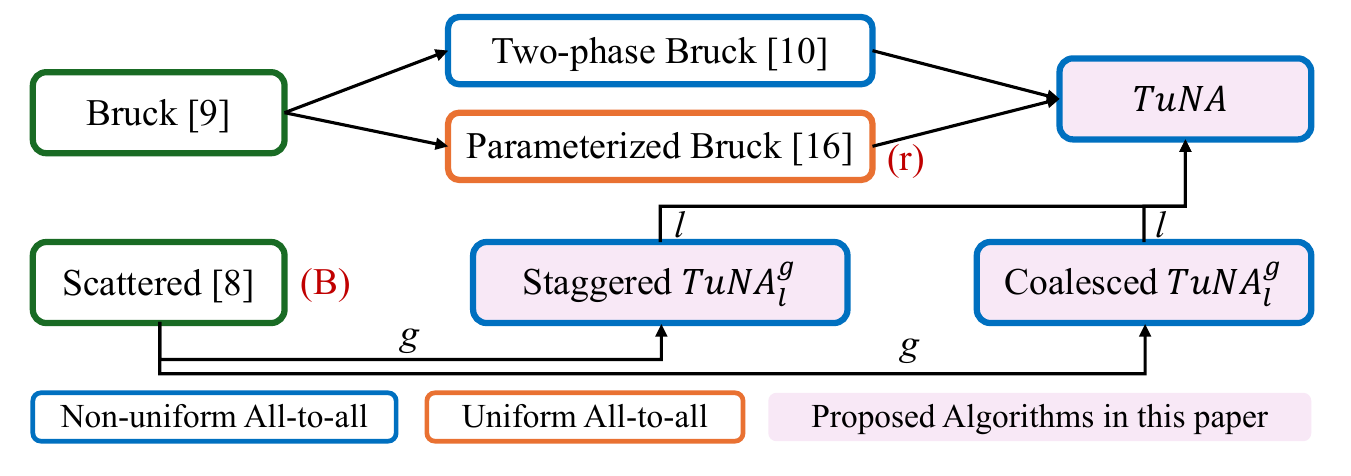} 
    \vspace{-1.5em}
    \caption{Interplay of proposed parameterized algorithms with existing foundational approaches.}
    \vspace{-1em}
    \label{fig:algsMap}
\end{figure}

\TRNA can be viewed as a synthesis of two algorithms: the two-phase \emph{non-uniform} Bruck~\cite{fan2022optimizing} and the parameterized \emph{uniform} Bruck~\cite{highradix2021,gainaru2016using,fan_configurable_2024}, both of which are derived from the classic Bruck algorithm. \TRNA, combined with the linear-time scattered algorithm, forms the foundation for the hierarchical \HTRNA{l}{g} algorithm. ~\autoref{fig:algsMap} illustrates the interplay between these base and derived algorithms. In the following sections, we provide a concise overview of the fundamental base algorithms to establish a clear context for our work.

\emph{(a) Bruck}~\cite{thakur2005optimization} is a classic logarithmic \emph{uniform} all-to-all algorithm with radix (base) $2$.
It comprises of three phases: an initial rotation phase, a communication phase with $log_2 P$ rounds, and an inverse rotation phase. 
It is a store-and-forward algorithm, where data-blocks received during one round are forwarded in subsequent rounds for further transmission.

\emph{(b) Two-phase non-uniform Bruck}~\cite{fan2022optimizing} is a logarithmic all-to-all algorithm with radix 2, designed for \emph{non-uniform} workloads.
The algorithm employs a coupled communication strategy consisting of metadata and data exchange phases to manage the \emph{non-uniform} workload.

\emph{(c) Parameterized Uniform Bruck} 
is the generalized version of the Bruck for \emph{uniform} workloads, wherein the radix can be tuned between 2 and $P$. This gives the ability to tune the number of communication rounds and the amount of data exchanged. 
Prior work~\cite{fan_configurable_2024} offers valuable insights into selecting the optimal radix, notably observing that \(r = \sqrt{P}\) yields the best overall performance.

\emph{(d) Standard non-uniform all-to-all implementations:} 
In both MPICH and OpenMPI,  \texttt{MPI\_Alltoallv} implementations employ variants of the spread-out (linear)~\cite{kang2020improving} algorithm. Spread-out schedules all send and receive requests in a round-robin order, ensuring that each process sends to a unique destination per round to avoid network congestion.
The \emph{scattered} algorithm in MPICH further improves this by dividing the communication requests into batches, wherein a tunable parameter, \emph{block-count}, decides the size of batches.
It waits for all the requests in one batch to be completed before moving on to the next, further reducing network congestion.
OpenMPI's linear approach deviates from spread out, as it initiates all communication in ascending rank order instead of round-robin. OpenMPI's other implementation, called \emph{pairwise} algorithm, initiates a single receiving request with the non-blocking \texttt{Irecv} and opts for a blocking \texttt{Send}. It then awaits the completion of these two requests per communication round.

%% file: trav.tex
\section{Tunable-radix Non-uniform All-to-all (\TRNA)}
\label{sec:trna}

\begin{figure}[tbp]
    \centering
    \vspace{-1.2em}
    \includegraphics[width=\linewidth, trim = 0.5cm 11.5cm 11cm 4.5cm, clip]{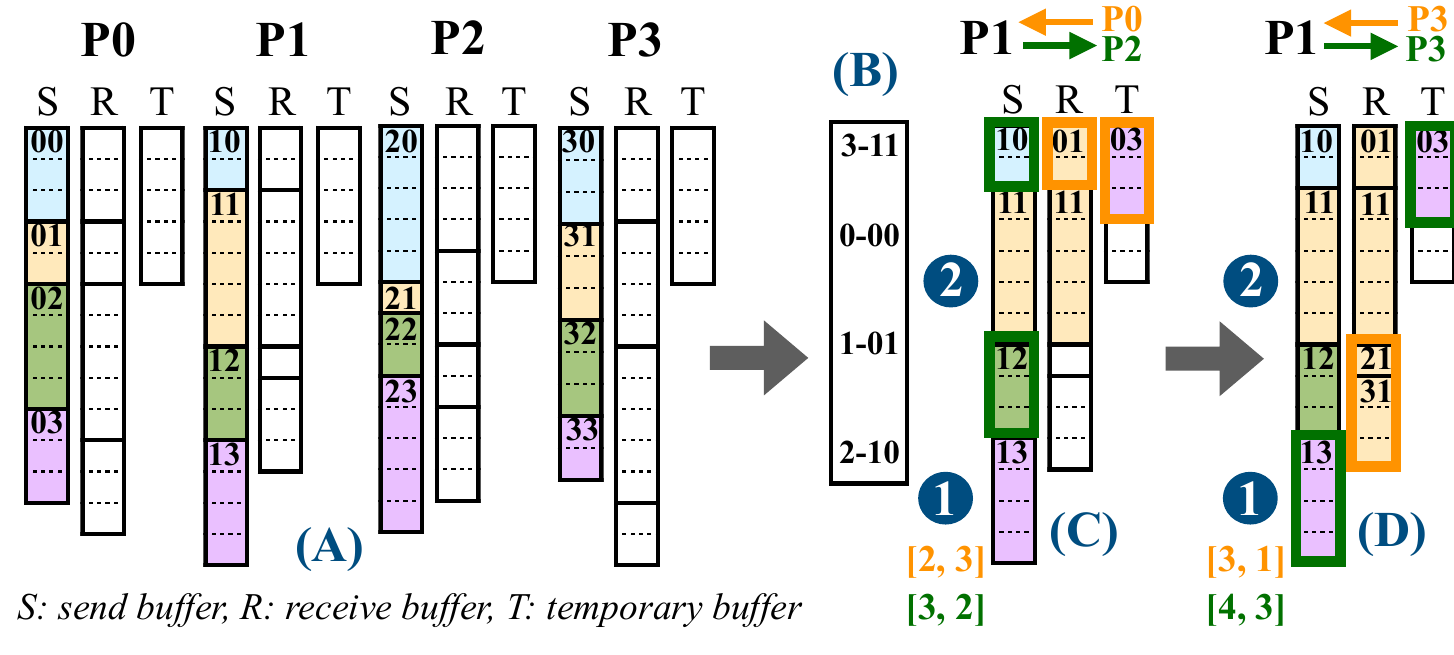}
    \vspace{-1.5em}
    \caption{Example of the \TRNA with $P = 4$ and $r = 2$. (A) is the initial state. $S$ is made of $4$ data-blocks (of different sizes), shown in different colors. 
    (B) shows the rotated data-block indices and their matching binary representation for $P\mathit{1}$. (C) and (D) illustrate two communication rounds for $P\mathit{1}$. 
    A two-phase communication scheme is employed in each round: \textcircled{\small 1} metadata exchange,
    and \textcircled{\small 2} actual data exchange.}
    \vspace{-1.5em}
    \label{fig:TRNA_exap}
\end{figure}

We present the \textbf{tu}nable-radix \textbf{n}on-uniform \textbf{a}ll-to-all (\TRNA) algorithm, which facilitates \emph{non-uniform} all-to-all data exchanges in a logarithmic number of communication rounds, parameterized with a tunable radix ($r$).
\TRNA is founded upon three ideas: (1) a logarithmic-time parameterized implementation of the Bruck algorithm with varying radices (based on parameterized \emph{uniform} bruck). 
(2) a two-phase data exchange scheme for each iteration of the Bruck algorithm, consisting of a metadata exchange followed by the actual data exchange (based on two-phase \emph{non-uniform} Bruck), (3) and a carefully sized temporary buffer ($T$) to facilitate intermediate data exchanges during the logarithmic communication rounds.

\vspace{-1em}
\subsection{Tunable Radix} 
\label{sec:trna:params}

All all-to-all implementations execute $K$ point-to-point communication rounds, exchanging a total of $D$ data blocks. In the Bruck, $K = \log_2P$, and the linear time spread-out $K = P$. Within \TRNA, both $K$ and $D$ are parameterized on the radix $r$ and the number of processes $P$
as described below.

Every process encodes the indices of their $P$ data blocks using a $r$-base representation.
In this encoding scheme, the maximum number of digits required is denoted by $w = \lceil log_r P \rceil$, and each digit can assume one of $r$ unique values. For example, in~\autoref{fig:TRNA_exap}, $w = log_2 4 = 2$, and each digit is $0$ or $1$. Consequently, a given data exchange round $k \ (0 \leq k < K)$ can be uniquely identified by two variables: $x \ (0 \leq x < w)$ and $z\ (1 \leq z < r)$. The variable $x$ represents the digit position within the $r$-base encoding, while $z$ corresponds to the specific value of the digit at that position. This leads to $K \leq w \cdot (r-1)$ communication rounds.
During each round, each process sends the data blocks whose $x_{th}$ digit matches the value $z$ to the process with a rank distance of $z \cdot r^x$. For instance, in~\autoref{fig:TRNA_exap} (C), \emph{P1} sends the data-blocks whose $0_{th}$ digit equals $1$ to \emph{P2} whose rank distance is $1 \cdot 2^0 = 1$. Each process transmits up to $r^{w-1}$ data blocks per round, bounding the total number of data-blocks exchange $D$ across all rounds to $w \cdot (r-1) \cdot r^{w-1}$. 


This communication strategy shows that both~$K$ and~$D$ are functions of $r$. These two parameters exhibit an inverse correlation, meaning an increase in $K$ corresponds to a decrease in~$D$, and vice versa. $K$ represents the latency-related metric while~$D$ is the bandwidth-related metric. As a result, by increasing $r$, the algorithm can effectively transition from a latency-bound regime (low latency) to a bandwidth-bound regime (high bandwidth). This trade-off between~$K$ and~$D$, facilitated by adjusting~$r$, provides a mechanism for tuning the communication performance.

\begin{figure}[t]
    \centering
    \vspace{-1em}
    \includegraphics[width=1\linewidth, trim = 0cm 15cm 8cm 2cm, clip]{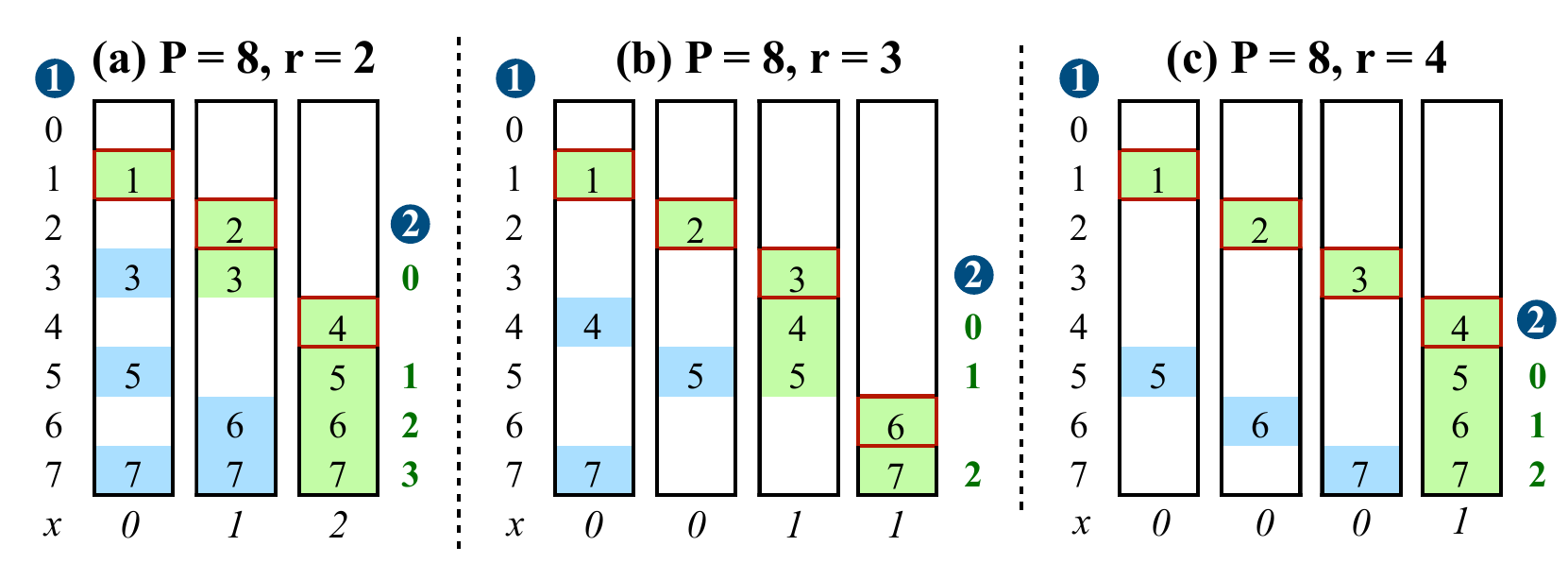}
    \vspace{-1.8em}
    \caption{Examples of memory optimization with three configurations, each showing a single process and the data blocks exchanged per communication round. In each round, green blocks reach their destination, while blue blocks are temporarily stored in $T$ for transfer in later rounds. Meanwhile, green blocks with red boxes are sent only once during the entire communication, allowing their space in $T$ to be omitted. }
    \vspace{-1.4em}
    \label{fig:temp-memory}
\end{figure}

\vspace{-1em}
\subsection{Two-phase communication} 
\label{sec:trna:twophase}

Although the parameterization allows us to adjust~$K$ and~$D$, it does not directly address the \emph{non-uniform} nature of the workload.
The parameterized Bruck is unsuitable for \emph{non-uniform} in its current form. This is because it employs a store-and-forward approach, where data blocks traverse multiple intermediate steps before reaching their final destination. Additionally, the algorithm utilizes send-and-receive buffers as temporary storage during intermediate communication phases.
To accommodate \emph{non-uniform} data distributions, we incorporate two key elements: a two-phase communication scheme and a temporary buffer ($T$). The scheme facilitates the exchange of intermediate data blocks, while $T$ provides the necessary storage for these blocks during the data exchange phase. 

The two-phase scheme is employed in each communication round; the first phase transfers the size of each sent data block, followed by the actual transmission of data. For instance, in~\autoref{fig:TRNA_exap} (C), process $P\mathit{1}$ needs to send data blocks $12$ and $10$ to process $P\mathit{2}$ (highlighted with green boxes). $P\mathit{1}$ first sends an array $[3, 2]$ to $P\mathit{2}$, representing the sizes of the two data blocks.
During the data exchange rounds, the sizes of the received data blocks may be larger than the sent ones in the send buffer ($S$) or the corresponding segments in the receive buffer~($R$). To solve this issue, the algorithm employs a temporary buffer~($T$) to accommodate for all intermediate received data blocks that will be transferred again in subsequent rounds, while data blocks destined for the current process are stored in $R$. 
For instance, in~\autoref{fig:TRNA_exap} (C), $P\mathit{1}$ receives data blocks $01$ in $R$ and $03$ in $T$ from $P\mathit{0}$.
In the next round (~\autoref{fig:TRNA_exap} (D)), $P\mathit{0}$ sends the data block from $T$ again to $P\mathit{3}$. 
Upon completing the communication phase, all processes receive the required data blocks in $R$, which lay in ascending order, as illustrated in~\autoref{fig:TRNA_exap} (D). Utilizing both $T$ and $R$, \TRNA effectively manages and rearranges the received data blocks, eliminating the overhead associated with the final rotation phase. 



\subsection{Estimating temporary buffer size} 
\label{sec:trna:mem}

Previous studies on modifying Bruck-like methods for \emph{non-uniform} workloads, including the two-phase non-uniform Bruck, adopted a specific approach to temporary buffer sizing~\cite{fan2022optimizing, xu2013sloavx}. They designed the temporary buffer ($T$) to accommodate all data blocks by setting its size to the product of two factors: the maximum block size ($M$) across all processes and the total number of processes ($P$).
While this approach works, it is wasteful in its memory requirements and can lead to memory overflow for large values $M$ or $P$. 
We observe that $T$ only needs to store intermediate data blocks, and therefore, a tighter bound on the size of $T$ can be obtained. To this end, we performed a theoretical analysis of the underlying communication pattern.

\begin{algorithm}[t]
\caption{\TRNA Algorithm}
\scriptsize
\begin{algorithmic}[1]
\label{code:trna}
\STATE Find maximum data-block size $M$ with \texttt{MPI\_Allreduce};
\STATE Allocate a temporary buffer $T$ with length $(N * (P - K - 1))$;
\STATE Compute $ti$ for each data-block $i$, $i \in [r+1, P]$.
\STATE Allocate rotation array $I$ for each process $p$;
\STATE $I[i] = (2 \times p - i + P) \ \% \ P, i \in [0, P]$;
\FOR{$x \in [0, w]$}
\FOR{$z \in [1, r]$}
\STATE $\emph{n} = 0$;
\FOR{$i \in [0, P]$ whose $x^\text{th}$ digit of \emph{r-base} is $1$}
\STATE $\emph{sd}[n\text{++}] = (p+i) \ \% \ P$   \qquad\qquad/* find $n$ send data-block indices */
\ENDFOR
\STATE $\emph{sendrank} = (p - z \times r^x + P) \ \% \ P$;
\STATE $\emph{recvrank} = (p + z \times r^x) \ \% \ P$;
\STATE Send metadata to \emph{sendrank} and receive updated metadata from \emph{recvrank};
\STATE Send sent data-block $I[\emph{sd}[i]]$ 
($i \in [0, n]$) to \emph{sendrank};
\IF{($i \% r^x == 0$) ($i \in [0, n]$)}
\STATE Receive data-block $i$ into $R$ from \emph{recvrank};
\ELSE
\STATE Receive data-block $i$ into $T$ from \emph{recvrank};
\ENDIF
\ENDFOR
\ENDFOR
\end{algorithmic}
\end{algorithm}


We observe that for every communication round, a process sends at most $r^x$ data blocks that reach their destination, while the remaining blocks are only transferred to some intermediate process that gets sent in coming communication rounds.
Here, $x$ refers to the digit of indices of data blocks in r-base encoding, ranging from $0$ to $w$ (see~\autoref{sec:trna:params}). For instance, ~\autoref{fig:temp-memory} (a) with $P = 8$ processes and $r = 2$ requires three rounds. In each round of this scenario, the data blocks needed to be temporarily stored in $T$ are marked in blue, while those reaching their destination are marked in green. For example, there are $2^1 = 2$ green data blocks in the second communication round.
Notably, we observed that in each round, the first data block is always sent directly to its destination process without further transfers in subsequent rounds, referred to as the \emph{direct} data block. 
In~\autoref{fig:temp-memory} (a), the green data blocks highlighted with red boxes indicate the \emph{direct} data block.
Since \emph{direct} data blocks do not need to be stored in the temporary buffer $T$, it allows us to put a tighter bound on its size. A process can set $T$ to store $B = (P - (K + 1))$ data-blocks, accounting for one block destined for itself.
We note that $(B)$ is a function of both $r$ and $P$, the value of which decreases as $r$ increases for a given $P$. For instance, in~\autoref{fig:temp-memory}, $r$ takes values of $2$, $3$, and $4$ in subfigures (a), (b), and (c), respectively. The corresponding value of $B$ for the three radices is $4$, $3$, and $3$. When $r$ exceeds $(P - 2)$, no temporary buffer is needed (equivalent to linear time spread out algorithm).

While our approach minimizes the size of $T$, it introduces the challenge of mapping the indices of data-block (referred to as $o$) into $T$. For example, in~\autoref{fig:temp-memory} (a), \textcircled{1} indicates the original indices ($o$) of data blocks, ranging from $0$ to $7$ and \textcircled{2} represents the corresponding mapped indices ($t$) in $T$, ranging from $0$ to $3$. Such as $o=3$ maps $t=0$ while $o=5$ maps $t=1$. 
The new position ($t$) is calculated as
$t = o - 1 - dx \cdot (r - 1) - dz$; this corresponds to taking the block's original index ($o$) and subtracting the number of direct data blocks with lower indices than the current block. 
Variables $dx$ and $dz$, mirrors $x$ and $z$ explained in Section~\ref{sec:trna:params}. $dx$ represents the highest digit when encoding index $o$ of a data block in the r-representation and equals $\lceil log_r o \rceil$, and $dz$ represents the value of that digit, which equals $o / r^dx$. 




\textbf{Algorithm:} The pseudocode of \TRNA is shown in Algorithm~\ref{code:trna}. It shows the three core ideas of \TRNA: (1) a logarithmic number of data exchange phases executed using a tunable number of communication rounds. The number of rounds is parameterized by the variables $x$ and $z$, as shown in lines $6$ and $7$.
(2) Each data exchange phase comprises two sub-phases: a meta-data exchange and the actual data exchange. This two-phase approach is evident in lines $14$ and $15$ of the pseudocode. (3) Creation and usage of an optimally sized temporary buffer ($T$), as shown in lines $2$ and $19$.

%% file: ltrav.tex
\begin{figure}[t]
    \centering
    \includegraphics[width=.85\linewidth, trim = 5cm 8cm 7.5cm 6.5cm, clip]{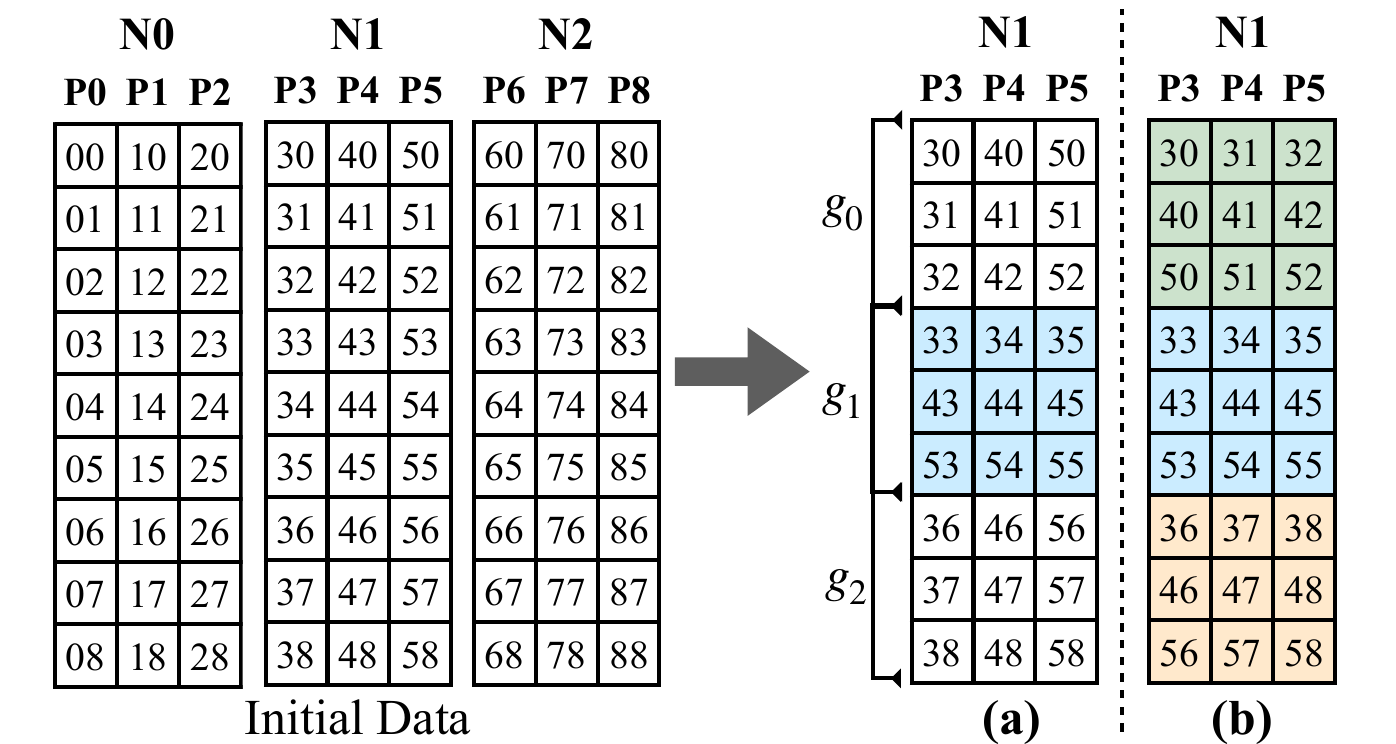}
    \vspace{-0.5em}
    \caption{Two intra-node strategies: (a) explicit and (b) implicit (ours). Assuming data blocks on each node are logically divided into $N=3$ groups, each process within a node has $Q=3$ data-blocks per group. An explicit strategy performs all-to-all only within the group whose index matches the node's ID. Our approach performs all-to-all within each group.}
    \vspace{-1.4em}
    \label{fig:intra_strag}
\end{figure}

\section{Hierarchical Tunable Non-uniform All-to-all}
\label{sec:hierach}

\begin{figure*}[tbp]
    \centering
    \vspace{-2em}
    \includegraphics[width=\textwidth, trim = 3cm 4.5cm 2cm 4.5cm, clip]{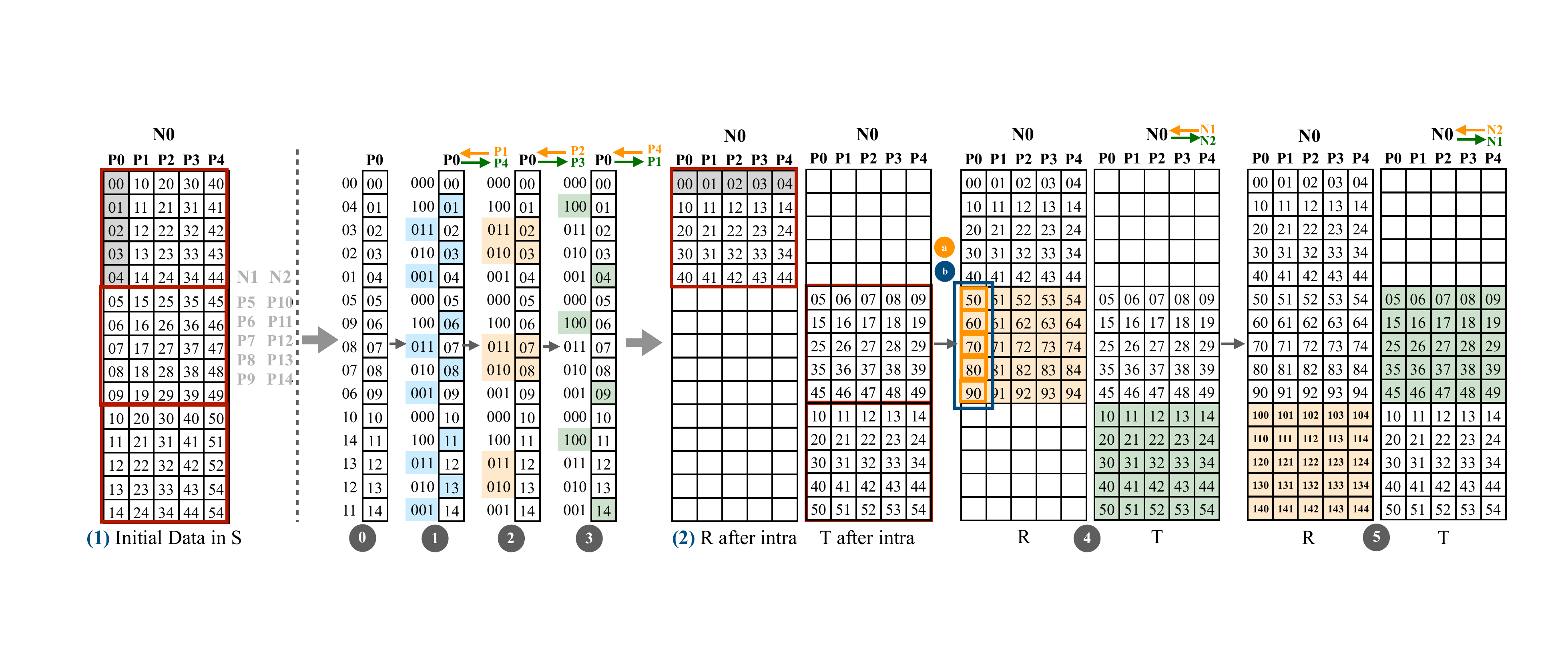}
    \vspace{-1.5em}
    \caption{An example of \HTRNA{l}{g} when $P = 15$, $N = 3$, $r = 2$ and $Q = 5$. (1) depicts the initial state in $S$ for all processes within node $N\mathit{0}$. Each process logically has $Q$ data-blocks in each $N$ group (separated by red boxes). \textcircled{\small 0} shows the rotation index array for $P\mathit{0}$ based on the group rank ID ( $g = p\ \%\ Q$ ). \textcircled{\small 1} \textcircled{\small 2} \textcircled{\small 3} illustrate the three intra-node communication steps, where the sent data-blocks are colored. (2) presents the data status in $R$ and $T$ after the intra-node communication, where $R$ holds the data-blocks destined the processes within $N\mathit{0}$. \textcircled{\small 4} \textcircled{\small 5} depict two communication steps for inter-node communication. 
    \textcircled{\small a} and \textcircled{\small b} are two communication patterns (matching~\autoref{fig:inter_strag}).
    Finally, each process receives the required data-blocks in $R$.
    }
    \vspace{-1em}
    \label{fig:trna-ts}
\end{figure*}


Modern HPC systems feature a hierarchical architecture, wherein each computing node comprises multiple CPU cores with shared memory access~\cite{jocksch2019optimized}.
This facilitates rapid intra-node data exchanges via direct memory transfers, typically significantly faster than inter-node exchanges (over the network).
To utilize the intra-node bandwidth, we present the hierarchical \textbf{tu}nable \textbf{n}on-uniform \textbf{a}ll-to-all (\HTRNA{l}{g}) algorithms (see Section~\ref{sec:hierach:HTRNA}). 
\HTRNA{l}{g} algorithms decouple the communication phase into two phases: (1) \textbf{l}ocal (or intra-node) communication and (2) \textbf{g}lobal (or inter-node) communication. The intra-node phase uses shared memory within a node, while the inter-node communication transfers messages over the network. 

Theoretically, \HTRNA{l}{g} algorithms can integrate any pair of \emph{non-uniform} all-to-all algorithms. 
\TRNA itself serves as an adaptable component with a tunable radix, accommodating both logarithmic and linear communication patterns.
However, \TRNA algorithms require a metadata exchange phase and a temporary buffer to manage \emph{non-uniform} data distributions appropriately. Utilizing it in both phases of \HTRNA{l}{g} might lead to high memory demands and inefficient data transfer between buffers. 
We thus only consider employing \TRNA in intra-node and scattered in the inter-node communication. Moreover, the inter-node data exchange deals with aggregated data, which works better with bandwidth-bound linear algorithms (for example, scattered). 
The scattered algorithm, adjustable in \emph{block\_count}, modulates the number of simultaneous network communication requests, further enabling performance optimization.
Additionally, we introduce two implementations of \HTRNA{l}{g} based on the distinct communication patterns of inter-node (see Section~\ref{sec:hierach:versions}): (1) staggered \HTRNA{l}{g} and (2) coalesced \HTRNA{l}{g}.
\subsection{Hierarchical \HTRNA{l}{g} Algorithm}
\label{sec:hierach:HTRNA}
\HTRNA{l}{g} is composed of intra-node and inter-node data exchanges, which we now present in detail.


\paragraph{Intra-node communication} 

With a total of $P = Q \cdot N$ processes, where $Q$ represents the number of processes per node and $N$ represents the total number of nodes, the intra-node data exchange phase consists of $N$ concurrent all-to-all exchanges, each of which involving $Q$ processes.
In standard all-to-all, each process ($p$) must send one data block ($i$) to process $i$ and receive one data-block ($p$) from process $i$, with the total number of data blocks equalling the number of processes ($P$). 
In our intra-node communication, all $P$ processes are logically grouped into $N$ groups (indexed $0,\ldots,N-1$), each containing $Q$ processes. Subsequently, we group the $P$ data blocks into $N$ groups, with every group handling $Q$ data-blocks.
We then perform all-to-all exchanges concurrently in each group using the \TRNA algorithm. ~\autoref{fig:intra_strag} (b) shows an example involving three nodes, where we perform three concurrent $Q \times Q$ all-to-all exchanges for each node. Our approach differs from an explicit approach, which creates a local sub-communicator for every group by splitting the MPI communicator using \texttt{MPI\_Comm\_split}. This approach only allows a local all-to-all exchange for the group whose index equals the node's ID ($n$) (see~\autoref{fig:intra_strag} (a)), as opposed to our approach, which allows concurrent all-to-all within every group. Our implicit approach avoids the overhead of creating new local communicators and also better prepares the data blocks for the subsequent inter-node data exchange.

We note that to achieve our implicit strategy, a meta-data exchange is required to exchange the data block sizes destined for processes in other nodes.
Fortunately, the intra-node communication phase of the \HTRNA{l}{g} algorithm employs the \TRNA algorithm, which internally includes a meta-data exchange phase, requiring no extra cost. Finally, we note that the intra-node communication phase in \HTRNA{l}{g} is implemented through \TRNA and is thus tunable with a radix $r$ $\in[0,\ldots,Q]$.


\begin{figure}[tbp]
    \centering
    \includegraphics[width=\linewidth, trim = 2.5cm 6.8cm 7cm 8cm, clip]{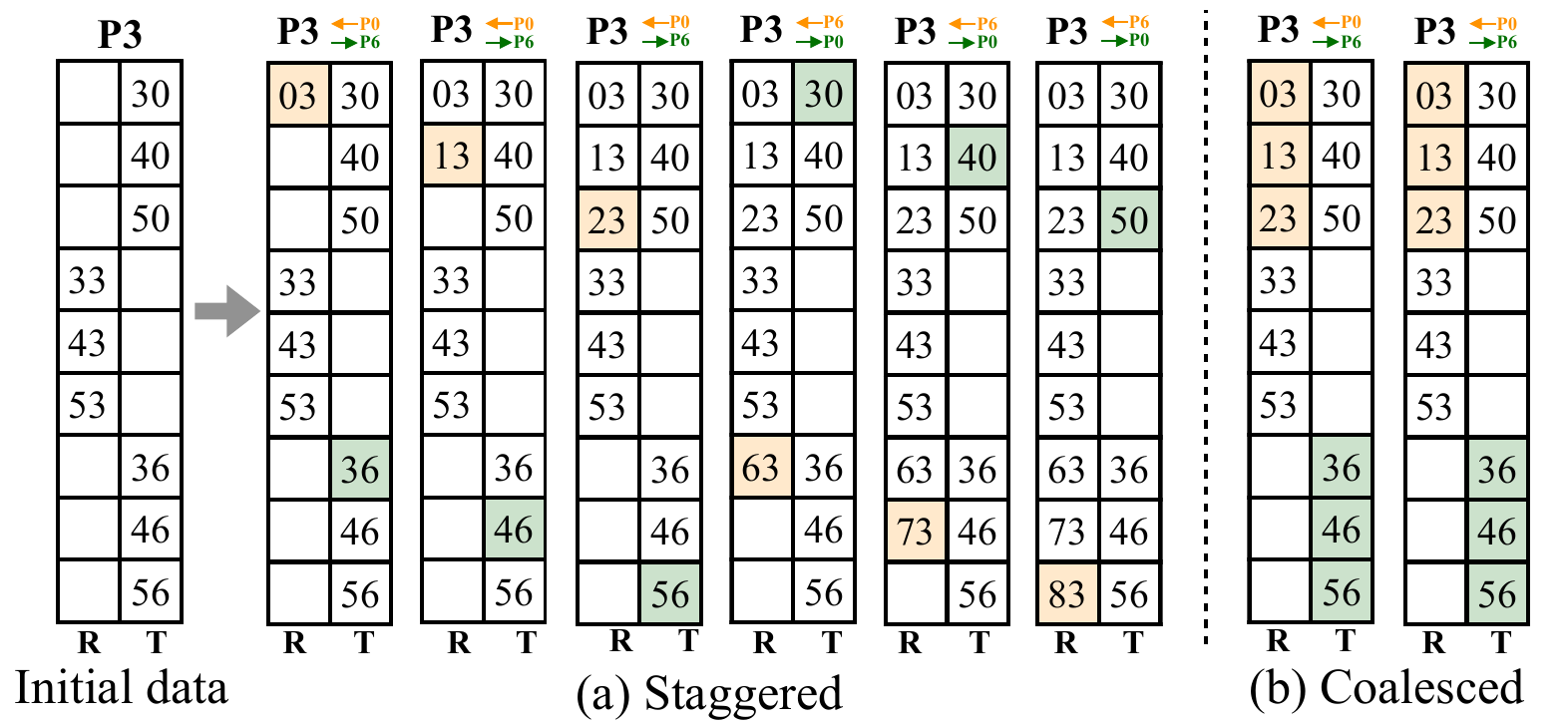}
    \vspace{-1.5em}
    \caption{Two inter-node communication patterns: (a) staggered and (b) coalesced.  Taking the first node $P\mathit{3}$ as an example, each process in (a) sends/receives one data-block to the same destination per round, requiring $(N - 1) \cdot Q$ rounds. In (b), each process sends/receives $Q$ data-blocks per round, requiring $(N - 1)$ rounds. See Section~\ref{sec:hierach:versions} for details.}
    \vspace{-1.5em}
    \label{fig:inter_strag}
\end{figure}

\paragraph{Inter-node communication} This phase conducts the all-to-all communication across nodes. In this phase, all $Q$ processes within a node must communicate with all $Q$ processes in another node. The communication process pairs have the same number of group ID $(g = p\ \%\ Q)$. This follows the $Q$-port model, in which every $Q$ point-to-point data exchange is delivered simultaneously. Each node serves as a communication port, transmitting a $\sfrac{1}{Q}$ message to the corresponding process in another node. 
For example, in~\autoref{fig:trna-ts} \textcircled{\small 4}, node $N\mathit{0}$ needs to send all green data-blocks in $T$ to node $N\mathit{2}$, while receiving all orange data-block in $R$ from node $N\mathit{1}$. In this case, each process sends its own $Q$ orange data-blocks to the matching process. 
The inter-node communication utilizes a scattered algorithm with an adjustable \emph{block\_count} to manage the communication load. This algorithm divides the communication requests into manageable batches, executed sequentially to mitigate network congestion. The \emph{block\_count} parameter determines the batch size, which can significantly influence the overall performance.

This inter-node communication is analogous to the concept of inter-group communication in MPI, implementable by using the \texttt{MPI\_Intercomm\_create} routine, creating an inter-communicator to define the local and remote groups. All processes in the local group exchange data with all processes in the remote group. 
However, MPI's inter-group approach is limited to only communication between exactly two groups, posing a constraint on scalability and flexibility. In contrast, we implement the inter-node exchange using point-to-point data exchanges, explicitly computing the ranks of all point-to-point exchange pairs, allowing our approach to support concurrent inter-communication among multiple groups.


\begin{algorithm}[tbp]
\caption{Staggered \HTRNATS Algorithm}
\scriptsize
\begin{algorithmic}[1]
\label{code:staggered}
\STATE Same intra-node communication with coalesced \HTRNATS Algorithm
\FOR{($ii = 0; ii < P; ii\ += \emph{block\_count}$)}
\FOR{$i \in [0, \emph{block\_count}]$}
\STATE $gi = (ii + i) / n$; $gr = (ii + i)\ \%\ n$; $nsrc = (g + gi)\ \%\ N$;
\IF{($nsrc\ != g$)}
\STATE $d = nsrc * Q + gr$; $src =  nsrc * n + g$;
\STATE Receive data-blocks $d$ from \emph{src} using \texttt{MPI\_Irecv};
\ENDIF
\ENDFOR
\FOR{$i \in [0, \emph{block\_count}]$}
\STATE $gi = (ii + i) / n$; $gr = (ii + i)\ \%\ n$; $ndst = (g - gi + N)\ \%\ N$;
\IF{($ndst\ != g$)}
\STATE $d = ndst * Q + gr$; $dst =  ndst * n + g$;
\STATE Send data-blocks $d$ to \emph{dst} using \texttt{MPI\_Isend};
\ENDIF
\ENDFOR
\STATE Wait for communication completion using \texttt{MPI\_Waitall}
\ENDFOR
\end{algorithmic}
\end{algorithm}
\subsection{Staggered and coalesced \HTRNA{l}{g}}
\label{sec:hierach:versions}

To further optimize the \HTRNA{l}{g} algorithm across diverse communication scenarios, we develop two variants based on distinct inter-node communication patterns (see~\autoref{fig:inter_strag}): (a) staggered \HTRNATS and (b) coalesced \HTRNATS.
The staggered communication pattern involves sequentially exchanging one data block per communication round with the target process, completing the communication within two nodes over $Q$ rounds (see~\autoref{fig:inter_strag} (a)). Inter-node communication requires $(N-1)$ exchanges between nodes. This method, therefore, requires a total of $Q\cdot(N - 1)$ communication rounds.
Conversely, the coalesced one consolidates the transmission, sending all $Q$ data blocks in a single round to the target process (see~\autoref{fig:inter_strag} (b)). This approach yields $(N - 1)$ communication rounds.
Theoretically, the staggered variant is effective for short-message scenarios, where the number of communication rounds dominates the performance. In contrast, the coalesced variant is more appropriate for bandwidth-bound long-message scenarios.

\textbf{Algorithms:} Algorithms~\ref{code:staggered} and ~\ref{code:coalesced} provide pseudocode for the staggered and coalesced \HTRNATS, respectively. Both share the same intra-node communication phase, detailed in lines $6$ to $18$ of Algorithm~\ref{code:coalesced}. The inter-node communication phases are outlined in lines $20$ to $30$ of Algorithm~\ref{code:coalesced} for the coalesced method and lines $2$ to $18$ of Algorithm~\ref{code:staggered} for the staggered method.
It is important to note that after the intra-node communication, noncontinuous data blocks are stored in buffer $T$. Hence, a local data rearrangement is necessary to eliminate any empty intermediate segments in $T$, thus streamlining the buffer for efficient coalesced inter-node communication. 

\begin{algorithm}[tbp]
\caption{Coalesced \HTRNATS Algorithm}
\scriptsize
\begin{algorithmic}[1]
\label{code:coalesced}
\STATE Find maximum global data-block length $M$ across processes;
\STATE Allocate a temporary buffer $T$ with length $(M * Q)$;
\STATE $g = p\ \%\ Q$; \quad $n = p\ /\ Q$; \quad $w = \lceil log_r Q \rceil$;
\STATE Allocate rotation array $I$ for each process $p$;
\STATE $I[i*Q + j] = i \times Q + (2 \times g - j + Q)\
\%\ Q$; $ i \in [0, N], j \in [0, Q]$;

\FOR{$x \in [0, w]$}
\FOR{$z \in [1, r]$}
\STATE $\emph{n} = 0$;
\FOR{$i \in [0, P]$ whose $x^\text{th}$ digit of \emph{r-base} is $1$} 
\STATE $\emph{sd}[n\text{++}] = n \times Q + (g + i) \ \% \ Q$   \qquad\qquad/* find $n$ send data-block indices */
\ENDFOR
\STATE $\emph{sendrank} = n \times Q + (g - z \times r^x + Q)\ \%\ Q$;
\STATE $\emph{recvrank} = n \times Q + (g + z \times r^x) \ \% \ Q$;
\STATE Send metadata to \emph{sendrank} and receive updated metadata from \emph{recvrank};
\STATE Send sent data-block $I[\emph{sd}[i]]$ 
($i \in [0, n]$) to \emph{sendrank};
\STATE Receive data-block $i$ into $T$ from \emph{recvrank};
\ENDFOR
\ENDFOR
\STATE Rearrange $T$ to removing empty data-blocks; 
\FOR{($ii = 0; ii < N; ii\ += \emph{block\_count}$)}
\FOR{$i \in [0, \emph{block\_count}]$}
\STATE $\emph{nsrc} = (n + i +ii)\ \%\ Q$; $\emph{src} = \emph{nsrc} \times Q + g$;
\STATE Receive data-blocks ranging from \emph{nsrc} to $(\emph{nsrc} + Q)$ from \emph{src};
\ENDFOR
\FOR{$i \in [0, \emph{block\_count}]$}
\STATE $\emph{ndst} = (n - i - ii + Q)\ \%\ Q$; $\emph{dst} = \emph{ndst} \times Q + g$;
\STATE Send data-blocks ranging from \emph{ndst} to $(\emph{ndst} + Q)$ to \emph{ndst};
\ENDFOR
\STATE Wait for communication completion using \texttt{MPI\_Waitall};
\ENDFOR
\end{algorithmic}
\end{algorithm}

%% file: evaluation.tex
\vspace{-0.3em}
\section{Evaluation}
\label{sec:eva}
\vspace{-0.2em}

We thoroughly evaluated our algorithms using micro-benchmarks on two production supercomputers: Polaris at Argonne National Laboratory and Fugaku at RIKEN R-CCS.
Polaris' $560$ nodes have $32$-core AMD CPUs and $4$ Nvidia A100 GPUs each, totaling a peak performance of $44$ petaflop/s. A Slingshot-based Dragonfly topology connects the nodes.
The $488$ Pflop/s \mbox{Fugaku} fields $158,\!976$ compute nodes, each hosting $48$ user-accessible A64FX cores. Fugaku's network is a 6D-torus \mbox{Tofu-D} interconnect.
Polaris uses Cray MPICH, while Fujitsu provides an OpenMPI-based version for Fugaku. 


To demonstrate the efficacy of \emph{\TRNA} and \emph{\HTRNA{l}{g}}, we evaluate their performance against vendor-optimized, closed-source implementations of \texttt{MPI\_Alltoallv} in Section~\ref{sec:eva:trna} and Section~\ref{sec:eva:HTRNA}. 
In addition to assessing the performance of standard \emph{non-uniform} all-to-all communication algorithms, we implement the four algorithms in OpenMPI and MPICH (detailed in Section~\ref{sec:background}).
Subsequently, we compare our algorithms, optimized with the best parameter configurations, against the top-performing \texttt{MPI\_Alltoallv} benchmark to demonstrate the efficiency of our approach (Section~\ref{sec:eva:comp}).

\begin{figure}[t]
    \centering
    \vspace{-1em}
    \includegraphics[width=\linewidth, trim = 3cm 3cm 2cm 2cm, clip]{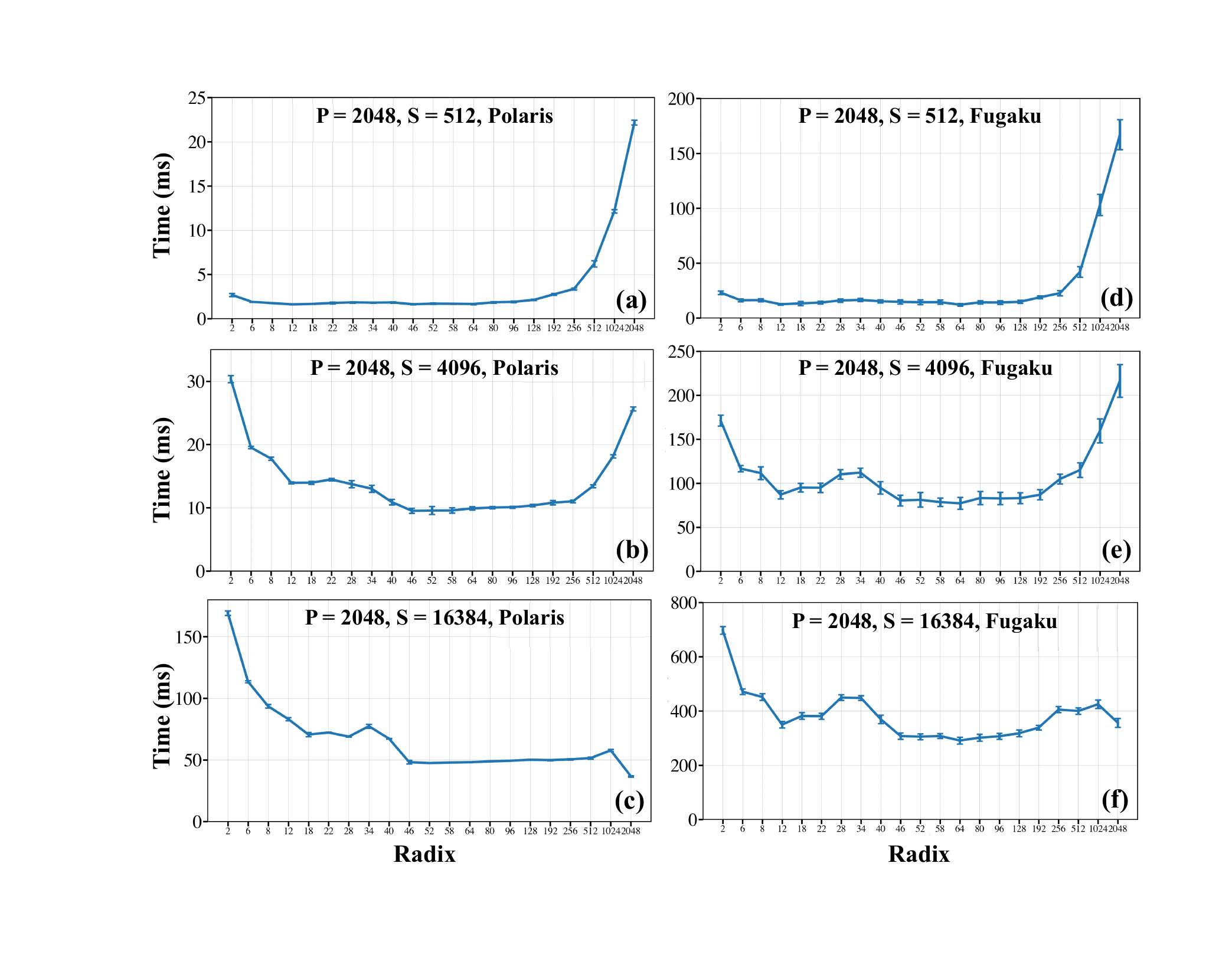}
    \vspace{-1.5em}
    \caption{Three trends of \TRNA on Polaris and Fugaku}
    \vspace{-1em}
    \label{fig:trna_tre}
\end{figure}

In all the aforementioned experiments, we vary the process count~($P$), radix~($r$), and maximum size of data blocks~($S$). 
By studying these parameters, we aim to determine optimal settings and evaluate how well our algorithms scale and adapt to various conditions.
All our experiments are performed for at least $20$ iterations, and we report the median and the standard deviations (using error bars). For Polaris and Fugaku, we utilized $32$ processes per node in our experiments.

\subsection{Performance analysis of \TRNA}
\label{sec:eva:trna}

In our experiments, every process generates data blocks whose sizes follow the continuous uniform distribution. 
This distribution ensures that data-block sizes are randomly selected and uniformly sampled between $0$ and $S$, thus yielding an average data block of size $\sfrac{S}{2}$.  
To understand the performance of \TRNA, we varied $S$ from $16$ bytes to \unit[16]{KiB} (generated using FP64 vectors), $r$ from $2$ to $P$, and $P$ from $512$ to $16,\!384$.

\textbf{Three performance trends:} Based on our experimental results, we identify three distinct performance trends for \TRNA when increasing radix $r$, which are consistent across all process counts ($P$): (1) for small $S$ ranging from $2$ to $512$ bytes, the performance of \TRNA exhibits an increasing trend with increasing radices. (2) For medium $S$ ranging from $512$ to \unit[8]{KiB}, the performance of \TRNA follows a \emph{U-shaped} trend. (3) For large $S$ exceeding \unit[8]{KiB} on Polaris and \unit[32]{KiB} on Fugaku, the performance of \TRNA shows a decreasing trend with increasing radix values. We can see these three trends in ~\autoref{fig:trna_tre} for $P = 2,\!048$.
Our measured ideal $r$ is around $2$ for the first trend involving small messages. This is attributed to small-sized message communication being dominated by latency, which requires minimal communication rounds to achieve optimal performance. 
The U-shaped trend suggests a balance between latency and bandwidth is sought for middle-sized message communication. Prior work has shown that $r \approx \sqrt{P}$ achieves this balance, minimizing the overall communication cost.
The bandwidth dominates the performance for large message communication, where the total transferred data size across all rounds becomes the critical factor. Therefore, $r \approx P$ is the sweet spot, as it minimizes the total transmitted message size.
Overall, the ideal $r$ increases when $S$ increases, transitioning from a latency-dominated regime to a bandwidth-dominated regime.



\begin{figure}[t]
    \centering
    \vspace{-1em}
    \includegraphics[width=\linewidth, trim = 3cm 2.5cm 2cm 12.5cm, clip]{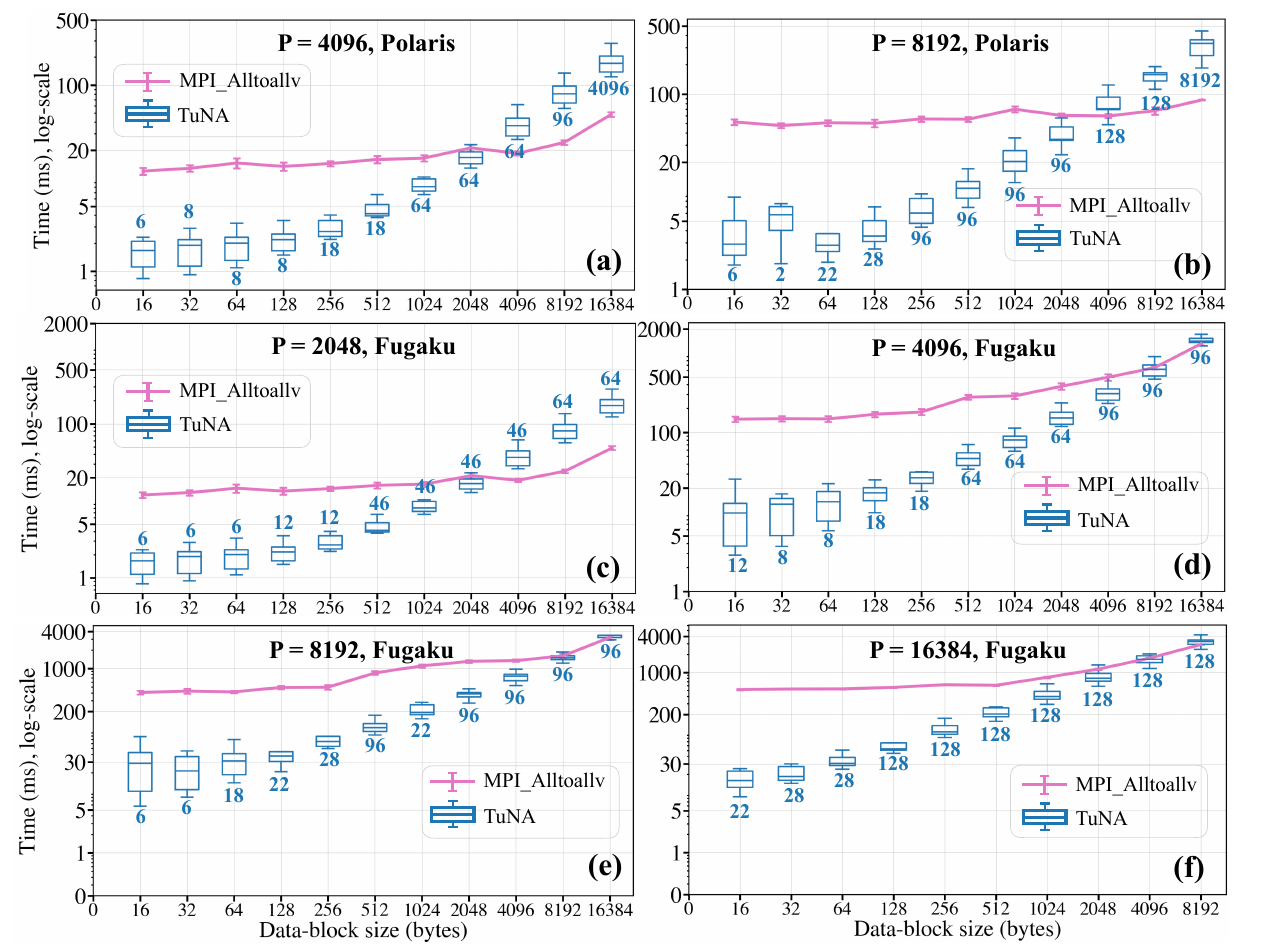}
    \vspace{-1.5em}
    \caption{Comparing \TRNA with \texttt{MPI\_Alltoallv} on Polaris and Fugaku. Detailed analysis provided in Section~\ref{sec:eva:trna}.}
    \vspace{-1.2em}
    \label{fig:trna_comp}
\end{figure}
\textbf{Performance comparison:} \autoref{fig:trna_comp} shows our comparisons of \TRNA against \texttt{MPI\_Alltoallv}.
The performance of \TRNA is presented through box plots, each representing the range of performance across various radices $\in [2,\ldots,P]$. 
Our measured ideal $r$ for each scenario is highlighted beneath its respective box, aligning with the above-observed trends of increasing $r$.
We see that \TRNA outperforms \texttt{MPI\_Alltoallv} when $S$ is no more than \unit[2]{KiB} on Polaris and \unit[16]{KiB} on Fugaku. 
Particularly, \TRNA demonstrates significant performance advantages when $S$ is less than \unit[512]{B} on Polaris and \unit[2]{KiB} on Fugaku.
For instance, when $P = 8,\!192$ and $S = 16$ bytes, \TRNA with ideal $r$ is $\sfrac{51.8}{1.78}=29\times$ and $\sfrac{408.33}{5.79}=70.48\times$ faster than \texttt{MPI\_Alltoallv} on Polaris and Fugaku, respectively.
\TRNA performs effectively with mid-ranged $S$. 
For example, it achieves $5.62\times$ and $7.26\times$ speedup on Polaris and Fugaku when $S = 1,\!024$ and $P = 8,\!192$.
While larger $S$ shows reduced performance on Polaris, \TRNA still manages a speedup of $9.69\times$ on Fugaku when $P = S = 16,\!384$.




\begin{figure}[t]
    \centering
    \vspace{-1em}
    \includegraphics[width=.90\linewidth, trim = 7cm 14.5cm 4cm 4cm, clip]{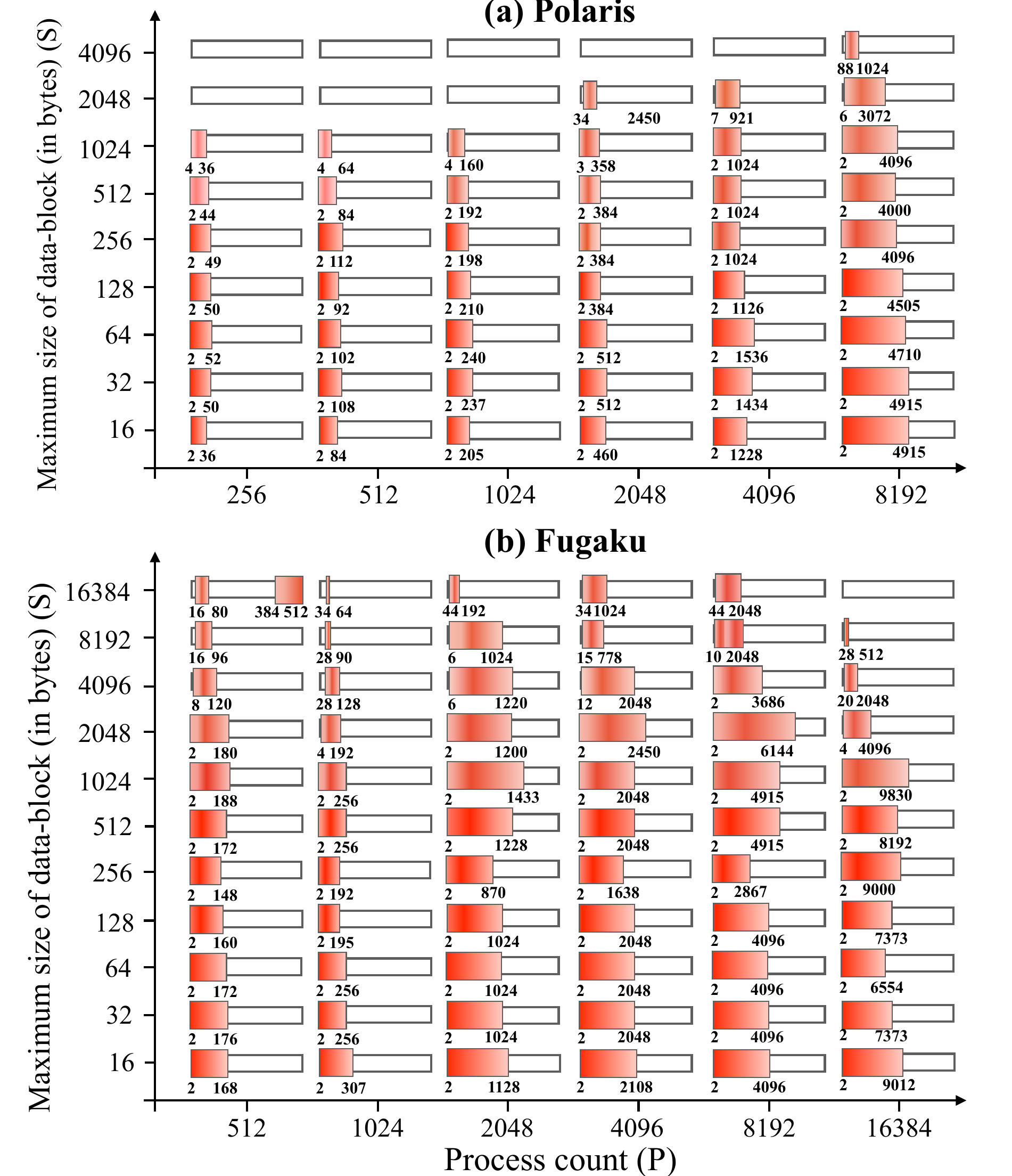}
    \vspace{-0.3em}
    \caption{Ranges of radix where \TRNA outperforms \texttt{MPI\_Alltoallv} on (a) Polaris and (b) Fugaku, visualized through a series of heatmaps. Each heatmap (top box) corresponds to a $P$ and $S$ pair, where the intensity of the red color indicates the degree of performance advantage offered by \TRNA. The bottom boxes indicate the entire radix range (from 2 to $P$). Refer to Section~\ref{sec:eva:trna} for details.} 
    \label{fig:trna_range}
    \vspace{-1em}
\end{figure}
\textbf{Radix selection of \TRNA:} We finally summarize all our experimental in \autoref{fig:trna_range}, which is developed to show the range of optimal radix of \TRNA that outperforms the vendor-optimized implementation of \texttt{MPI\_Alltoallv}. In this figure, $P$ is represented on the x-axis and $S$ on the y-axis. A combined shape containing two rectangles is presented for all combinations of $P$ and $S$.
The longer rectangle represents the entire range of radices for \TRNA, ranging from $2$ to $P$. The shorter rectangle, nested within the longer one, represents the specific range of radices for which \TRNA outperforms \texttt{MPI\_Alltoallv}.
Additionally, the shorter rectangle is depicted as a heatmap, where the intensity of the red color indicates the degree of performance improvement achieved by \TRNA. A stronger red color signifies a higher performance gain offered by \TRNA.
The previously mentioned trends can be observed from the heatmap, reinforcing the relationship between the optimal radix selection, $P$, and $S$.
These figures provide a concise and visually intuitive representation of the performance landscape, enabling us to make informed decisions when selecting suitable radices of \TRNA for given $P$ and $S$.

\subsection{Performance analysis of coalesced and staggered \HTRNA{l}{g}}
\label{sec:eva:HTRNA}

\begin{figure}[t]
    \centering
    \vspace{-1.2em}
    \includegraphics[width=\linewidth, trim = 7cm 2.5cm 2.1cm 2.5cm, clip]{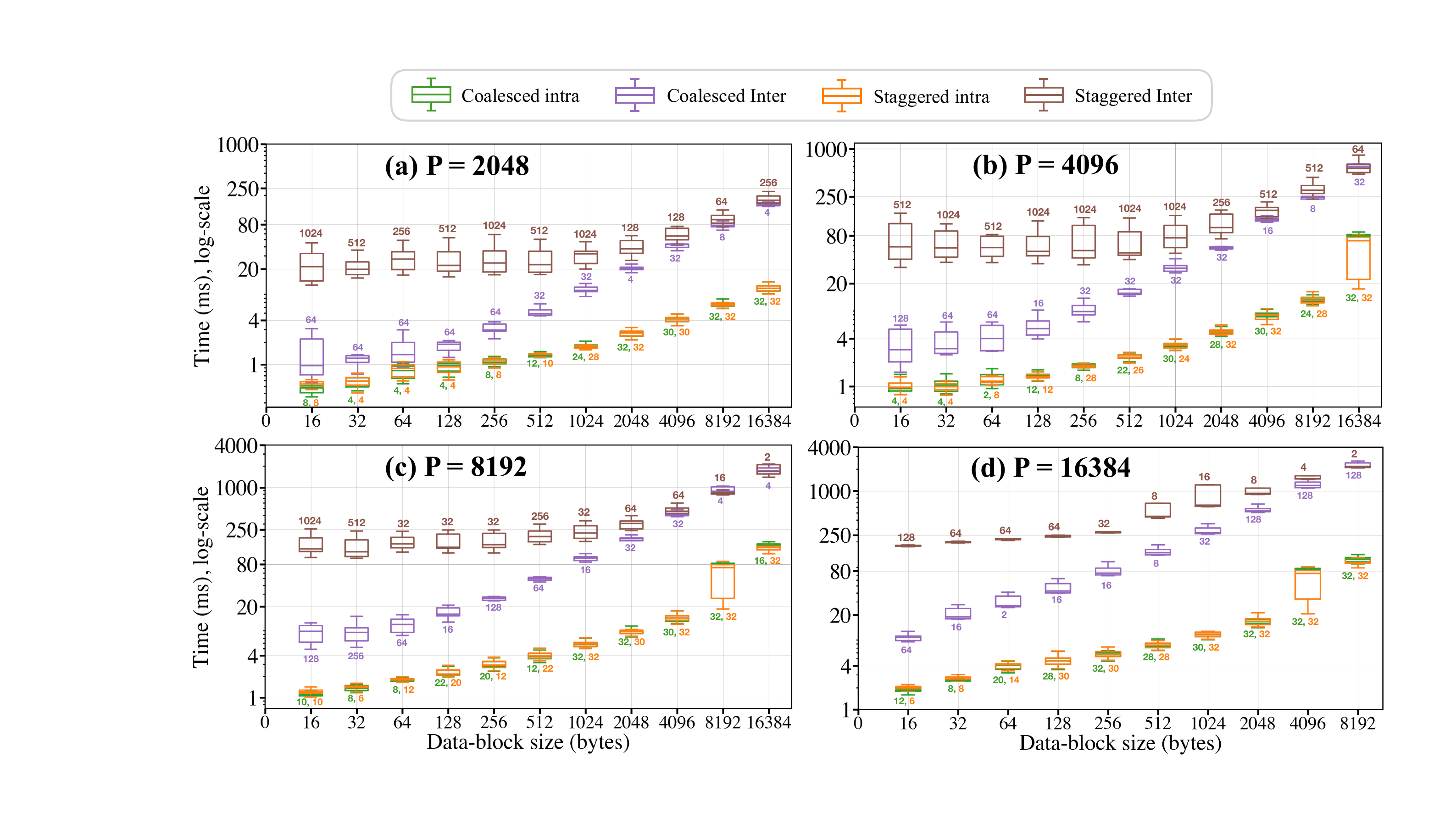}
    \caption{Comparing coalesced and staggered \HTRNA{l}{g} algorithms on Fugaku. Intra-node and inter-node data exchanges are plotted separately using box plots for each algorithm.}
    \label{fig:two-htrna}
\end{figure}

Both coalesced and staggered \HTRNA{l}{g} employ the \TRNA algorithm for intra-node communication and the scattered algorithm for inter-node communication. 
The \TRNA algorithm includes a configurable parameter, radix ($r$), which can be adjusted from $2$ to $Q$. Similarly, the scattered algorithm features a tunable parameter, \emph{block\_size}, which varies from 1 to $(N - 1)$ for the coalesced variant and from 1 to $((N-1) \cdot Q)$ for the staggered variant. 
\autoref{fig:two-htrna} presents the performance of these algorithms on Fugaku using box plots. In the experiments, the number of processors ($P$) was varied from $2,\!048$ to $16,\!384$, and the message size ($S$) from $16$ to \unit[16]{KiB}. 
Each box plot, distinguished by unique colors, represents either the intra-node or inter-node communication performance with its corresponding tunable parameter. 
Green and purple boxes depict the intra-node and inter-node communications of the staggered variant, while orange and brown boxes illustrate these phases for the coalesced one.
The ideal parameters for each configuration are indicated beneath each corresponding box plot with a matching color.

\begin{figure}[t]
    \centering
    \includegraphics[width=\linewidth, trim = 3cm 13.3cm 2.1cm 2.8cm, clip]{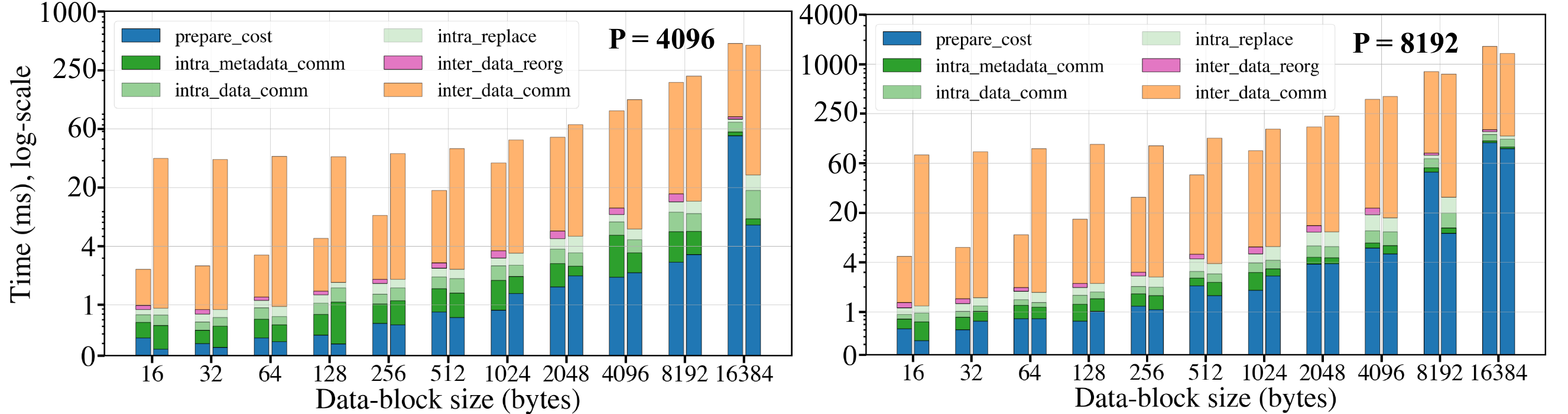}
    \vspace{-1em}
    \caption{Breakdowns of coalesced (left bar) and staggered (right bar) \HTRNA{l}{g} algorithms on Fugaku.}
     \vspace{-1.5em}
    \label{fig:breakdown}
\end{figure}
\vspace{-0.2em}
\textbf{Parameter selection analysis:} 
From~\autoref{fig:two-htrna}, we observe that the choice of optimal radix for intra-node communication does not follow a strict pattern; however, smaller radices generally perform better for smaller $S$ (under \unit[1]{KiB}), while larger radices are better for larger $S$. The observed pattern is reasonable because communication involving small message sizes tends to be more affected by the total number of communication rounds compared to communication with larger message sizes.
For inter-node communication, varying the \emph{block\_count} ($B$) shows a clear trend with increasing $S$, where larger $S$ typically favors smaller $B$.
For example, at $P = 8,\!192$, the ideal $B$ for staggered \HTRNA{l}{g} (indicated by the brown box) is $1,\!024$ and $2$ at $S = 16$ and \unit[16]{KiB}, respectively. 
Moreover, as $P$ increases, the ideal $B$ for the same $S$ tends to decrease. For example, with $S = 512$, the ideal $B$ for $P = 4,\!096$, $8,\!192$, and $16,\!384$ is $1,\!024$, $256$, and $8$, respectively. 
The choice of \emph{block\_count} significantly impacts performance, particularly for the staggered variant, which makes more communication requests than the coalesced configuration. 

\begin{figure}[t]
    \centering
    \vspace{-1.2em}
    \includegraphics[width=\linewidth, trim = 7cm 2.5cm 2.1cm 3cm, clip]{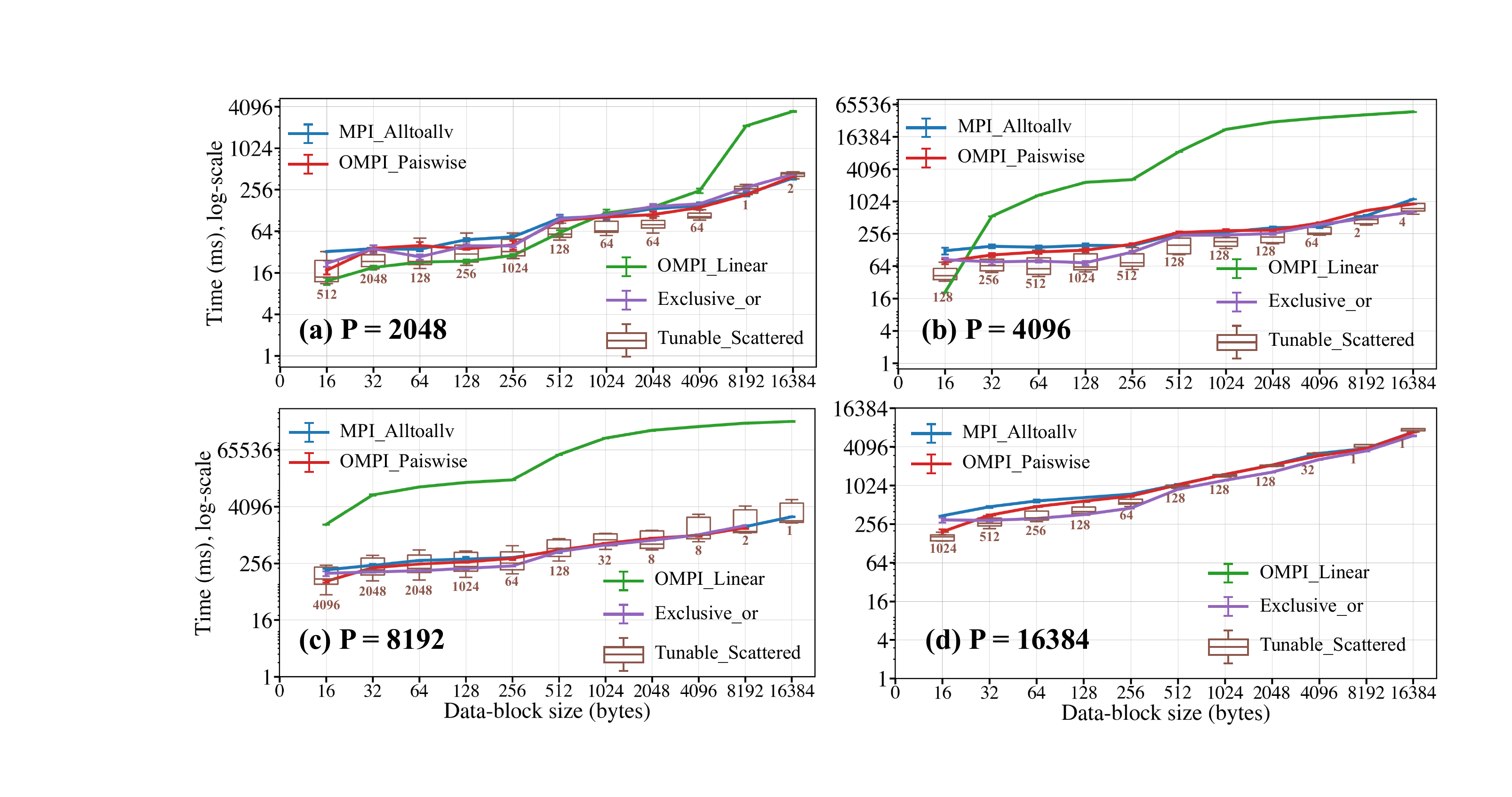}
    \caption{Benchmarking the non-uniform all-to-all implementations in OpenMPI and MPICH on Fugaku.}
    \vspace{-1.2em}
    \label{fig:mpi_exps}
\end{figure}

\begin{figure*}[t]
    \centering
    \vspace{-2em}
    \includegraphics[width=\linewidth, trim = 2.5cm 2.5cm 2cm 2cm, clip]{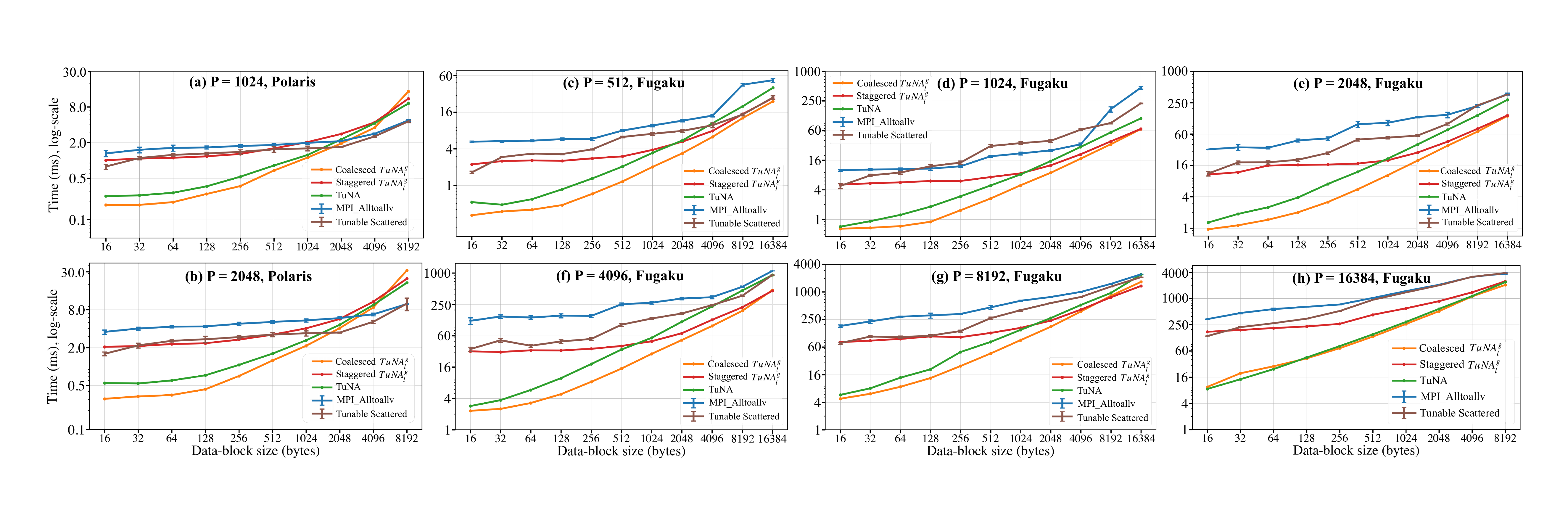}
    \vspace{-1.5em}
    \caption{Comparing the proposed algorithms with the top-performing benchmarks. Refer to Section~\ref{sec:eva:comp} for analysis.}
    \vspace{-1em}
    \label{fig:finalcomp}
\end{figure*}

\textbf{Performance comparison:} We also observe that the coalesced \HTRNA{l}{g} significantly outperforms the staggered one, particularly for small message sizes $S$. 
The staggered algorithm only exhibits competitive performance when $S$ is at least \unit[8]{KiB}.
For instance, at $P = 8,\!192$ and $S = 16$, the coalesced is $17.06\times$ faster than the staggered. Conversely, at $S = 16,\!384$ bytes, the staggered achieves a $1.23\times$ speedup compared to the coalesced.
\autoref{fig:breakdown} provides a detailed breakdown of the coalesced (left-bar) and staggered (right-bar) algorithms. 
The algorithms are divided into six components: (1) prepare-cost, which includes all preparatory steps (see lines $1$-$5$ and $9$-$13$ in Algorithm~\ref{code:coalesced}); (2) and (3) correspond to metadata-cost (line $14$) and data-cost (line $15$), respectively; (4) replace-cost, which covers the cost of inter-data copying in each round (line $16$); (5) data-rearrange time (line $19$), applicable only to the coalesced algorithm; (6) inter-node comm cost (lines $20$-$30$). 
The figure illustrates that the inter-node comm cost for the staggered is significantly higher than that for the coalesced.



\vspace{-0.2em}
\subsection{Comparison with vendor-optimized MPI}
\label{sec:eva:comp}

To evaluate the efficiency of our proposed algorithms, we benchmarked the performance of four standard \emph{non-uniform} all-to-all algorithms from MPI libraries (detailed in Section~\ref{sec:background}).
\autoref{fig:mpi_exps} presents a comparison of the default \texttt{MPI\_Alltoallv} on Fugaku and Polaris. Meanwhile, the performance of the scattered algorithm with a tunable \emph{block\_count} ($B$) is displayed using a box plot. We observe that OpenMPI's linear algorithm, which is a blocking linear algorithm, performs the worst, particularly for larger $P$. The pairwise, exclusive-or, and default \texttt{MPI\_Alltoallv} algorithms exhibit similar performance. Notably, when configured with ideal $B$, the scattered algorithm outperforms the others in most scenarios.

\textbf{Performance comparison:} Consequently, we compared our proposed algorithms, involving \TRNA, coalesced and staggered \HTRNA{l}{g}, against both the scattered algorithm and the default \texttt{MPI\_Alltoallv}. Each algorithm was configured with its ideal parameter.
\autoref{fig:finalcomp} illustrates this comparison, where all proposed algorithms surpass the default \texttt{MPI\_Alltoallv} across all scenarios and outperform the scattered algorithm in most instances.
Specifically, the performance improvements with small message sizes $S$ were notable, achieving maximum $60.60\times$ (\TRNA), $138.59\times$ (coalesced), and $12.29\times$ (staggered) speedups compared to \texttt{MPI\_Alltoallv} on Fugaku. 
Among these, the coalesced \HTRNA{l}{g} consistently demonstrated the highest performance across all scenarios. 
For instance, at $P = 16,\!384$ on Fugaku, it is $42.08\times$ and $14.61\times$ faster than \texttt{MPI\_Alltoallv} and the scattered algorithm at $S = 16$; and it maintained $2.20\times$ and $2.14\times$ improvements at $S = 8,\!192$.
Additionally, although our proposed algorithms perform suboptimally with large $S$ (larger than \unit[2]{KiB}), the coalesced one achieves $11.68\times$ and $2.71\times$ speedup at $ S = 64$ and \unit[2]{KiB}.

%% file: application.tex
\section{Applications}
\label{sec:app}

In this section, we assess the performance of our algorithms using real-world applications, including Fast Fourier Transform (see Section~\ref{sec:app:fft}), path-finding (see Section~\ref{sec:app:tc}), and two standard distributions (see Section~\ref{sec:app:dist}).

\vspace{-0.4em}
\subsection{Fast Fourier transform}
\label{sec:app:fft}
Fast Fourier Transform (FFT) computations are crucial for a wide range of scientific domains, such as fluid dynamics and astrophysics~\cite{cooley1965algorithm}.
Parallel FFT is characterized by performing three matrix transposes using all-to-all exchanges.
FFW3, an open-source parallel FFT library, distributes the sub-problems evenly among processes. When the problem size $\mathcal{N}$ is not an integer multiple of $P^2$, a \emph{non-uniform} all-to-all exchange is employed.
The data type of $\mathcal{N}$ is \texttt{fftw\_complex}, comprising two FP64 values representing a complex number's real and imaginary components. 

\begin{figure}[tbp]
    \centering
    \includegraphics[width=\linewidth, trim = 35.7cm 10.5cm 2.2cm 3cm, clip]{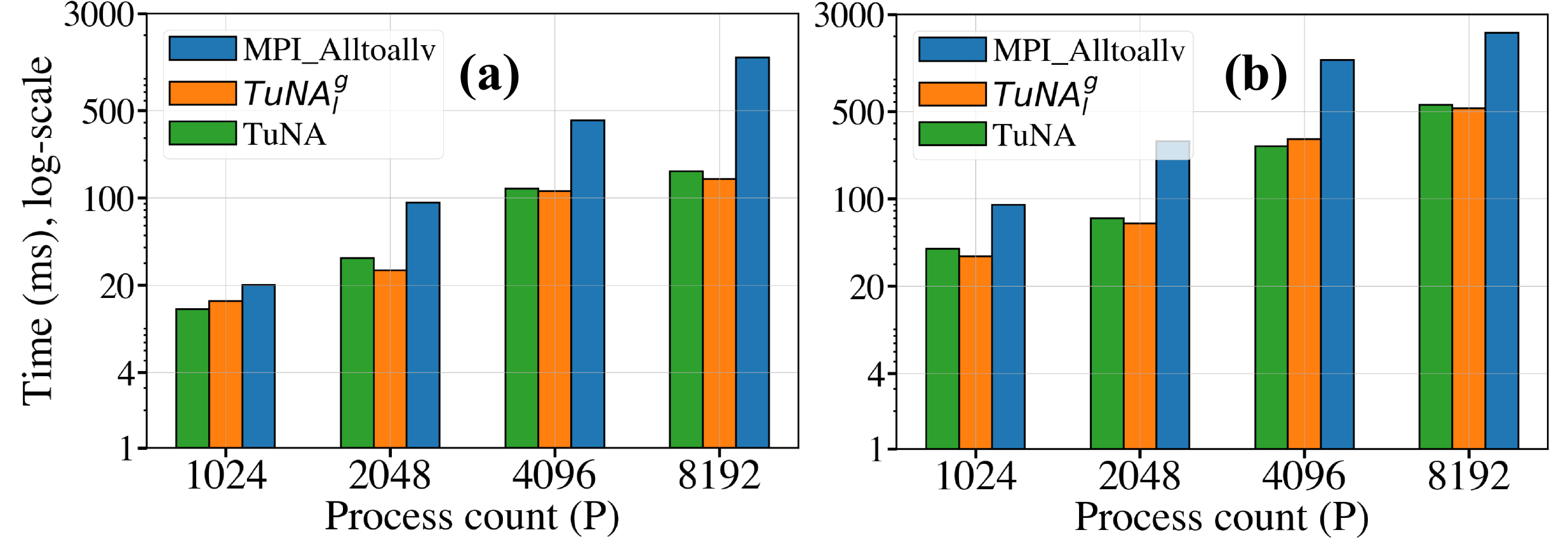}
    \vspace{-1.7em}
    \caption{Performance of applying our algorithms to an FFT HPC workload with two different input sizes $\mathcal{N}_1$ and $\mathcal{N}_2$.}
     \vspace{-1em}
    \label{fig:fft_app}
\end{figure}

We perform two experiments to test the effectiveness of our algorithms, featuring different \emph{non-uniform} data distributions using two values of $\mathcal{N}$: (1) $\mathcal{N}_1 = \lceil 0.78125 \cdot P \rceil \cdot \lceil P \cdot 0.625 \rceil \cdot 8$. This setup ensures that processes with ranks lower than $\lceil P \cdot 0.625 \rceil$ (referred to as \emph{worker}) are assigned data, while the remaining ranks receive no data. Each \emph{worker} fills the first $\lceil P \cdot 0.78125 \rceil$ data blocks with $8$ FP64 values.
(2) $\mathcal{N}_2 = ((P - 1) \cdot 32 + 8) \cdot P$. This leads to a near-uniform distribution where each process (except the last) transmits $64$ FP64 values, and the last one transmits $16$ FP64 values.

\autoref{fig:fft_app} presents the comparative results of our algorithms against \texttt{MPI\_Alltoallv} using the above mentioned two configurations
on Fugaku. We report the application runtime for the three approaches, dominated by all-to-all exchanges. 
For both experimental setups, all our algorithms outperform \texttt{MPI\_Alltoallv} for all process counts ($P$). Consistent with our observations made in Section~\ref{sec:eva:comp}, the \HTRNA{l}{g} (coalesced) steadily outperforms the other algorithms.
Additionally, our proposed algorithms demonstrate better performance for $\mathcal{N}_1$, which involves a smaller problem size. For example, when $P = 8,\!192$, the \HTRNA{l}{g} (coalesced) is $9.42\times$ and $4.01\times$ faster than \texttt{MPI\_Alltoallv} for $\mathcal{N}_1$ and $\mathcal{N}_2$, respectively.

\vspace{-0.2em}
\subsection{Graph Mining: path finding}
\label{sec:app:tc}
\vspace{-0.1em}

\begin{figure}[tbp]
    \vspace{-0.6em}
    \centering
    \includegraphics[width=\linewidth, trim = 2.2cm 11cm 2.2cm 1.5cm, clip]{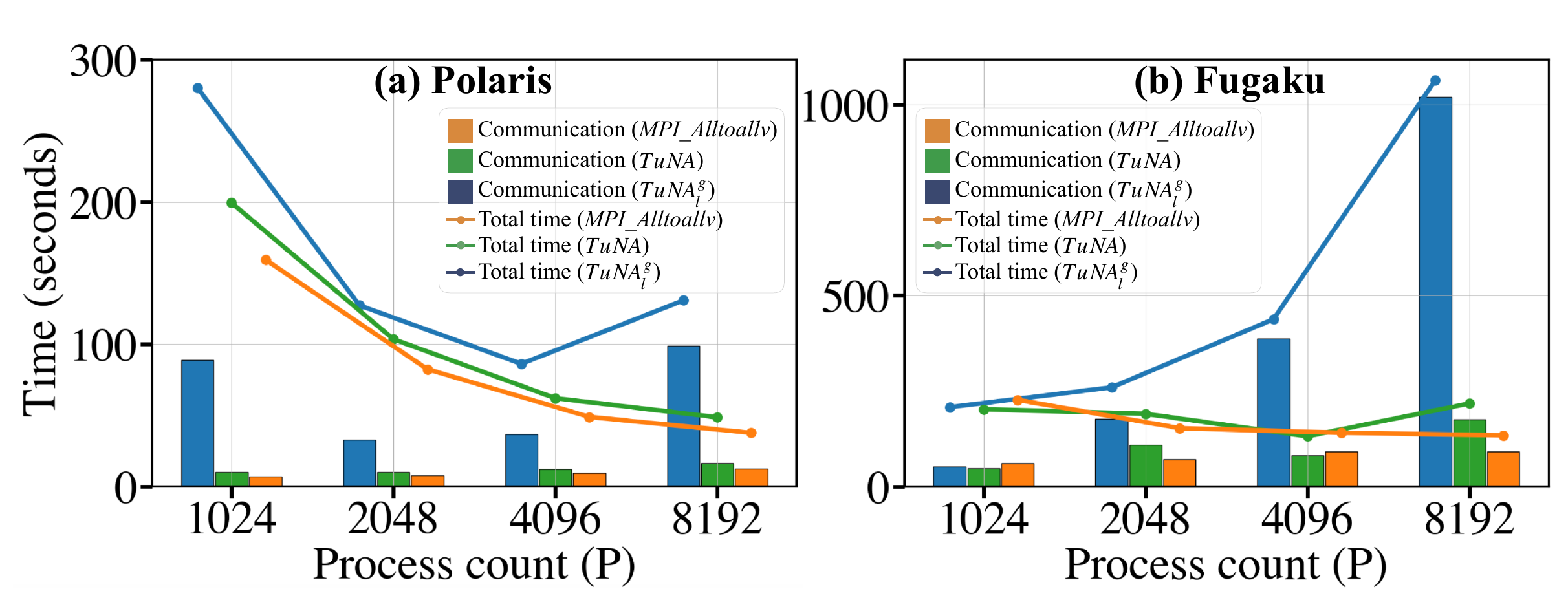}
    \vspace{-1.7em}
    \caption{Performance of applying our algorithms to path finding.}
     \vspace{-1em}
    \label{fig:tc}
\end{figure}

We evaluate our algorithms by applying them to a popular graph mining algorithm that computes all reachable paths in a graph, also known as the transitive closure of a graph~\cite{patel2021scalable, moustafa2016datalography}. 
The transitive closure (TC) of a graph can be computed through a classic fixed-point algorithm. that repeatedly applies a relational algebra (RA) kernel to a graph G. This operation discovers paths of increasing length within the graph. The process continues until no new paths can be identified, reaching the fixed point.
We use the MPI-based open-source library for parallel relational algebra~\cite{kumar:hipc:2019, gilray:cc:2021, fan2021exploring, kumar:isc:2020}, which utilizes \texttt{MPI\_Alltoallv} in each iteration of the fixed-point loop to shuffle data.
Our proposed algorithms maintain the same function signature as \texttt{MPI\_Alltoallv}, and hence, they can be seamlessly substituted in its place. 
We use a graph with $1,\!014,\!951$ edges, sourced from the Suite Sparse Matrix Collection~\cite{Davis:2011:UFS:2049662.2049663}. This graph undergoes repeated all-to-all across more than $5,\!800$ iterations to reach the fixed point. 

\autoref{fig:tc} presents the strong scaling performance comparison of our proposed algorithms, configured ideally as described in Section~\ref{sec:eva}, against the vendor-optimized \texttt{MPI\_Alltoallv}, using the same graph on Polaris (a) and Fugaku (b). 
In both subfigures, we depict the communication overhead with bar charts and the total execution time with line charts. These figures highlight the critical role of all-to-all communication in this application. The results demonstrate that our proposed algorithms outperform \texttt{MPI\_Alltoallv} in most cases. For instance, at $P = 8,\!192$, \TRNA achieves speedups of $5.98\times$ on Polaris and $5.80\times$ on Fugaku, while \HTRNA{l}{g} (coalesced) delivers even greater improvements of $7.96\times$ and $11.09\times$, respectively. Additionally, \HTRNA{l}{g} (coalesced) generally surpasses \TRNA in performance, with the exceptions being cases that could be further optimized through parameter tuning. These observations align with those discussed in Section~\ref{sec:eva}. It is worth noting that, despite the application’s limited scalability, our proposed algorithms still provide significant performance gains over \texttt{MPI\_Alltoallv} on both machines.


\vspace{-0.3em}
\subsection{Standard distributions}
\label{sec:app:dist}

In addition to the applications previously discussed and the uniform distribution presented in Section~\ref{sec:eva}, we further assess the effectiveness and generalization of our algorithms by validating them against two standard distributions: a power-law (exponential) distribution and a normal (Gaussian) distribution. 
~\autoref{fig:stand-dists} (a) and (b) depict the communication data for process 0 with $P = 4,\!096$ on Fugaku, following the two distributions, with a maximum data block size of 1024 bytes.

\textbf{\emph{Normal:}} ~\autoref{fig:stand-dists} (c) presents a weak scaling performance comparison of our algorithms against \texttt{MPI\_Alltoallv} on Fugaku, with workload size following a normal distribution. The results show that all of our algorithms outperform \texttt{MPI\_Alltoallv}, with coalesced \HTRNA{l}{g} demonstrating the best performance in almost all cases. In contrast, staggered \HTRNA{l}{g} performs worse than the other proposed two algorithms. For instance, at $P = 4,\!096$, \TRNA, coalesced \HTRNA{l}{g}, and staggered \HTRNA{l}{g} are $3.21\times$, $3.63\times$, and $1.57\times$ faster than \texttt{MPI\_Alltoallv}, respectively. This observation aligns with the results reported in Section~\ref{sec:eva} with the uniform distribution.

\textbf{\emph{Power-law:}} Similarly, ~\autoref{fig:stand-dists} (d) shows results for the power-law workload distribution, which is characterized by the rarity of large-sized data blocks and the sparsity of the data distribution. From the figure, we can draw conclusions similar to those made for the uniform and normal distributions. Notably, both \TRNA and coalesced \HTRNA{l}{g} significantly outperform \texttt{MPI\_Alltoallv}, particularly at large scales. 

%% file: related_work.tex
\section{Related Work}
\label{sec:relWork}

While substantial research efforts~\cite{faraj2005automatic, venkata2012exploring, traff2014implementing} went into optimizing all-to-all for \emph{uniform} messages, the exploration of \emph{non-uniform} data-loads has received comparatively little attention. Most relevant to our work are studies~\cite{xu2013sloavx, fan2022optimizing}, which adapted Bruck's algorithm (with a radix of 2) for \emph{non-uniform} all-to-all. 

\begin{figure}[h]
    \centering
    \includegraphics[width=\linewidth,  trim = 11.5cm 2.7cm 2.2cm 10cm, clip]{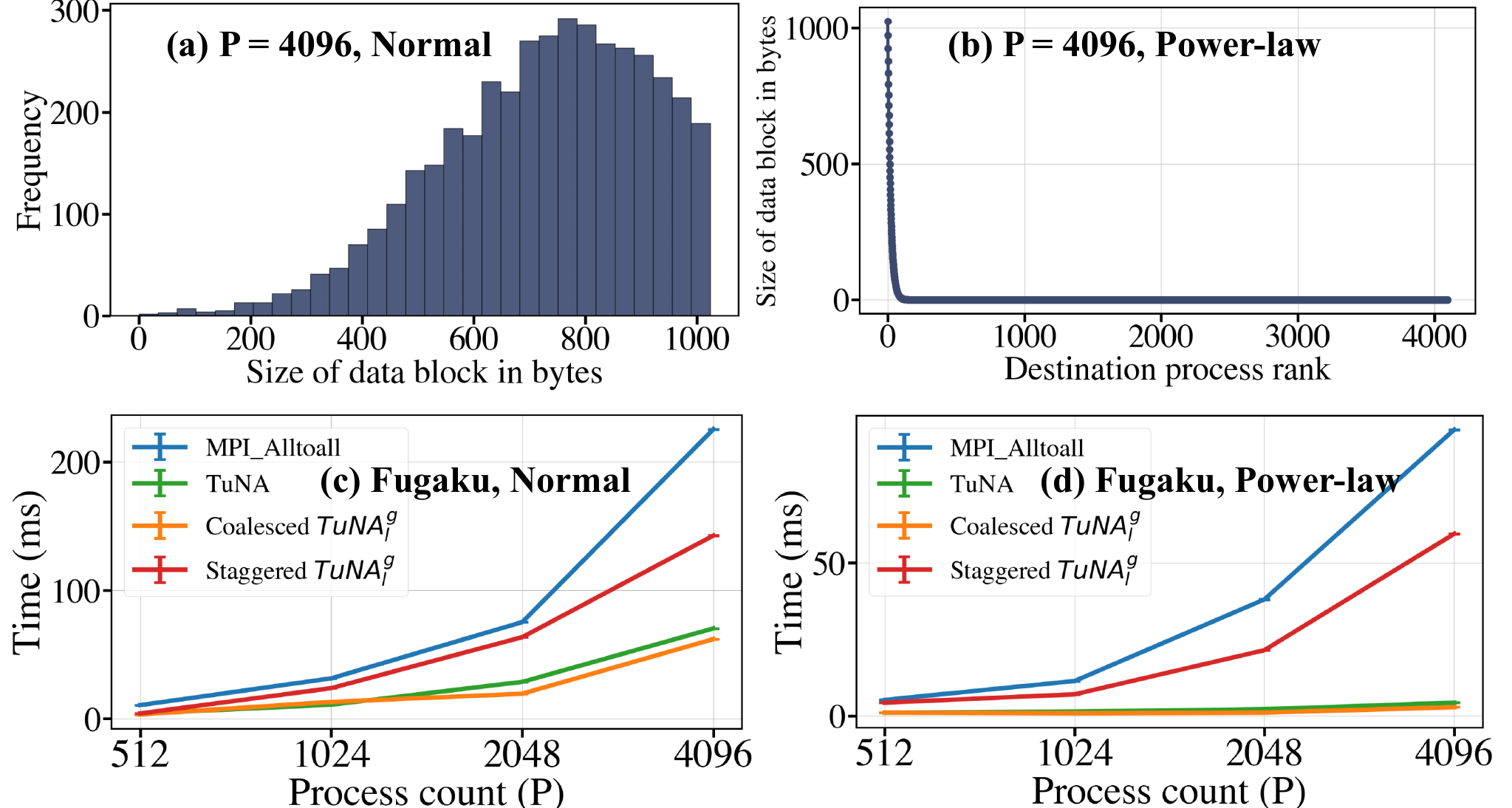}
    \vspace{-1.2em}
    \caption{(a) histogram to show normal data distribution used (mean: $1,\!000$, and standard deviation: $240$) and (b) shows power-law data distribution (exponent: $0.95$); (c) and (d) present a performance comparison for these two distributions.}
    \vspace{-1.4em}
    \label{fig:stand-dists}
\end{figure}

\textbf{Bruck variants:}
Tr{\"a}ff et al.~\cite{traff2014implementing} presented two improvements over Bruck, termed modified Bruck and zero-copy Bruck. The former omits the final rotation phase by rearranging data blocks in an initial rotation. The latter aims at reducing internal memory duplications.
Cong et al.~\cite{xu2013sloavx} eliminates the shifting of data-blocks during the initial rotation by employing an index array that stores the desired order of data blocks, effectively avoiding the actual movement of data.
Subsequently, Fan et al.~\cite{fan2022optimizing} further tuned Bruck by refine~\cite{traff2014implementing} and~\cite{xu2013sloavx} to eliminate the initial and final rotation phase.


\textbf{Tunable radix-based collectives:}
Gainaru et al.~\cite{gainaru2016using} studied logarithmic all-to-all algorithms with varying radices, but focused on exploring various memory layout configurations.
Taru et al.~\cite{highradix2021} presented all-to-all with high radices. However, it does not conduct performance analysis of varying radices nor provide a heuristic for selecting radices.
Andreas et al.~\cite{jocksch2020optimised} investigated the efficiency of collectives, with emphasis on \emph{allgatherv}, \emph{reduce-scatter}, and \emph{allreduce} with varying radices. Their research indicates that high radices work for  shorter messages and low radices work for longer ones.


\textbf{Hierarchical collectives:}
Jackson et al.~\cite{jackson2004planned} introduced a planned \emph{non-uniform} all-to-all, which transmits data from all processes on the same node to a designated master. Subsequently, only the master participate in a global exchange, reducing network congestion and hence demonstrating gains for concise messages.
Similarly, Plummer et al.~\cite{plummer2004lpar} segmented all processes into non-intersecting groups. Within each group, processes transfer data to the leading process,
which then participates in the (sparse) all-to-all, improving scenarios where data load distribution is irregular but remains constant over time.
Bienz et al.~\cite{bienz2022locality} proposed a locality-aware allgather built upon the Bruck algorithm, which clusters processes into groups of regions exhibiting low communication overhead.

%% file: conclusion.tex
\section{Conclusion}

We tackle the complex problem of optimizing the performance of \emph{non-uniform} all-to-all exchanges by proposing two novel algorithms, \TRNA and \HTRNA{l}{g}. \TRNA is a parameterizable algorithm that outperforms vendor-optimized MPI implementations on two production supercomputers. Building upon \TRNA, \HTRNA{l}{g} adds a hierarchical design to leverage the fast memory buffers of modern systems. Splitting the communication into node-local and global components further improves efficiency. Our experiments with up to 16k MPI ranks on up to 512 compute nodes show that \HTRNA{l}{g} outperforms both \TRNA and the vendor-provided \texttt{MPI\_alltoallv}. To showcase the effectiveness of our algorithms, we apply them to real-world applications, gaining nearly $10\times$ speedup.
In summary, our techniques can improve the performance of a wide range of applications relying on \emph{non-uniform} all-to-all. Applications and vendors can easily adopt our open-source implementations, offering an interface equivalent to \texttt{MPI\_Alltoallv} paired with tunable parameters for optimal performance.

%% file: main.bbl
\begin{thebibliography}{10}
\providecommand{\url}[1]{#1}
\csname url@samestyle\endcsname
\providecommand{\newblock}{\relax}
\providecommand{\bibinfo}[2]{#2}
\providecommand{\BIBentrySTDinterwordspacing}{\spaceskip=0pt\relax}
\providecommand{\BIBentryALTinterwordstretchfactor}{4}
\providecommand{\BIBentryALTinterwordspacing}{\spaceskip=\fontdimen2\font plus
\BIBentryALTinterwordstretchfactor\fontdimen3\font minus \fontdimen4\font\relax}
\providecommand{\BIBforeignlanguage}[2]{{%
\expandafter\ifx\csname l@#1\endcsname\relax
\typeout{** WARNING: IEEEtran.bst: No hyphenation pattern has been}%
\typeout{** loaded for the language `#1'. Using the pattern for}%
\typeout{** the default language instead.}%
\else
\language=\csname l@#1\endcsname
\fi
#2}}
\providecommand{\BIBdecl}{\relax}
\BIBdecl

\bibitem{bernholdt2020survey}
D.~E. Bernholdt, S.~Boehm, G.~Bosilca, M.~Gorentla~Venkata, R.~E. Grant, T.~Naughton, H.~P. Pritchard, M.~Schulz, and G.~R. Vallee, ``A survey of mpi usage in the us exascale computing project,'' \emph{Concurrency and Computation: Practice and Experience}, vol.~32, no.~3, p. e4851, 2020.

\bibitem{chen_highly_2022}
C.-C. Chen, K.~S. Khorassani, Q.~G. Anthony, A.~Shafi, H.~Subramoni, and D.~K. Panda, ``Highly {{Efficient Alltoall}} and {{Alltoallv Communication Algorithms}} for {{GPU Systems}},'' in \emph{2022 {{IEEE International Parallel}} and {{Distributed Processing Symposium Workshops}} ({{IPDPSW}})}, May 2022, pp. 24--33.

\bibitem{zhou_accelerating_2024}
Q.~Zhou, B.~Ramesh, A.~Shafi, M.~Abduljabbar, H.~Subramoni, and D.~Panda, ``Accelerating {{MPI AllReduce Communication}} with {{Efficient GPU-Based Compression Schemes}} on {{Modern GPU Clusters}},'' in \emph{{{ISC HIGH PERFORMANCE}} 2024}, May 2024.

\bibitem{besta_push_2017}
M.~Besta, M.~Podstawski, L.~Groner, E.~Solomonik, and T.~Hoefler, ``To {{Push}} or {{To Pull}}: {{On Reducing Communication}} and {{Synchronization}} in {{Graph Computations}},'' in \emph{Proceedings of the 26th {{International Symposium}} on {{High-Performance Parallel}} and {{Distributed Computing}}}, ser. {{HPDC}} '17.\hskip 1em plus 0.5em minus 0.4em\relax {New York, NY, USA}: {Association for Computing Machinery}, 2017, pp. 93--104.

\bibitem{nvidia_corporation_multinode_2022}
{NVIDIA Corporation}, ``Multinode {{Multi-GPU}}: {{Using NVIDIA cuFFTMp FFTs}} at {{Scale}},'' \url{https://developer.nvidia.com/blog/multinode-multi-gpu-using-nvidia-cufftmp-ffts-at-scale/}, Jan. 2022.

\bibitem{willsch_large-scale_2023}
D.~Willsch, M.~Willsch, F.~Jin, H.~De~Raedt, and K.~Michielsen, ``Large-{{Scale Simulation}} of {{Shor}}'s {{Quantum Factoring Algorithm}},'' \emph{Mathematics}, vol.~11, no.~19, 2023.

\bibitem{collom_optimizing_2023}
G.~Collom, R.~P. Li, and A.~Bienz, ``Optimizing {{Irregular Communication}} with {{Neighborhood Collectives}} and {{Locality-Aware Parallelism}},'' in \emph{Proceedings of the {{SC}} '23 {{Workshops}} of {{The International Conference}} on {{High Performance Computing}}, {{Network}}, {{Storage}}, and {{Analysis}}}, ser. {{SC-W}} '23.\hskip 1em plus 0.5em minus 0.4em\relax {New York, NY, USA}: {Association for Computing Machinery}, 2023, pp. 427--437.

\bibitem{mpich-web}
https://www.mpich.org, {MPICH} Home Page.

\bibitem{thakur2005optimization}
R.~Thakur, R.~Rabenseifner, and W.~Gropp, ``Optimization of collective communication operations in mpich,'' \emph{The International Journal of High Performance Computing Applications}, vol.~19, no.~1, pp. 49--66, 2005.

\bibitem{fan2022optimizing}
K.~Fan, T.~Gilray, V.~Pascucci, X.~Huang, K.~Micinski, and S.~Kumar, ``Optimizing the bruck algorithm for non-uniform all-to-all communication,'' in \emph{Proceedings of the 31st International Symposium on High-Performance Parallel and Distributed Computing}, 2022, pp. 172--184.

\bibitem{domke_a64fx_2021}
J.~Domke, ``{{A64FX}} -- {{Your Compiler You Must Decide}}!'' in \emph{2021 {{IEEE International Conference}} on {{Cluster Computing}} ({{CLUSTER}}), {{EAHPC Workshop}}}.\hskip 1em plus 0.5em minus 0.4em\relax {Portland, Oregon, USA}: {IEEE Computer Society}, 2021.

\bibitem{sato_co-design_2020}
M.~Sato, Y.~Ishikawa, H.~Tomita, Y.~Kodama, T.~Odajima, M.~Tsuji, H.~Yashiro, M.~Aoki, N.~Shida, I.~Miyoshi, K.~Hirai, A.~Furuya, A.~Asato, K.~Morita, and T.~Shimizu, ``Co-{{Design}} for {{A64FX Manycore Processor}} and "{{Fugaku}}",'' in \emph{Proceedings of the {{International Conference}} for {{High Performance Computing}}, {{Networking}}, {{Storage}} and {{Analysis}}}, ser. {{SC}} '20.\hskip 1em plus 0.5em minus 0.4em\relax {Atlanta, GA, USA}: {IEEE Press}, 2020, pp. 1--15.

\bibitem{argonne_national_laboratory_polaris_2024}
{Argonne National Laboratory}, ``Polaris {\textbar} {{Argonne Leadership Computing Facility}},'' \url{https://www.alcf.anl.gov/polaris}, 2024.

\bibitem{highradix2021}
D.~Taru, N.~Islam, G.~Zheng, R.~Kalidas, A.~Langer, and M.~Garzaran, ``High radix collective algorithms,'' \emph{Proceedings of EuroMPI 2021}, 2021.

\bibitem{gainaru2016using}
A.~Gainaru, R.~L. Graham, A.~Polyakov, and G.~Shainer, ``Using infiniband hardware gather-scatter capabilities to optimize mpi all-to-all,'' in \emph{Proceedings of the 23rd European MPI Users' Meeting}, 2016.

\bibitem{fan_configurable_2024}
K.~Fan, S.~Petruzza, T.~Gilray, and S.~Kumar, ``Configurable {{Algorithms}} for {{All-to-all Collectives}},'' in \emph{{{ISC HIGH PERFORMANCE}} 2024}, May 2024.

\bibitem{kang2020improving}
Q.~Kang, R.~Ross, R.~Latham, S.~Lee, A.~Agrawal, A.~Choudhary, and W.-k. Liao, ``Improving all-to-many personalized communication in two-phase i/o,'' in \emph{SC20: International Conference for High Performance Computing, Networking, Storage and Analysis}.\hskip 1em plus 0.5em minus 0.4em\relax IEEE, 2020, pp. 1--13.

\bibitem{xu2013sloavx}
C.~Xu, M.~G. Venkata, R.~L. Graham, Y.~Wang, Z.~Liu, and W.~Yu, ``Sloavx: Scalable logarithmic alltoallv algorithm for hierarchical multicore systems,'' in \emph{2013 13th IEEE/ACM International Symposium on Cluster, Cloud, and Grid Computing}.\hskip 1em plus 0.5em minus 0.4em\relax IEEE, 2013, pp. 369--376.

\bibitem{jocksch2019optimized}
A.~Jocksch, M.~Kraushaar, and D.~Daverio, ``Optimized all-to-all communication on multicore architectures applied to ffts with pencil decomposition,'' \emph{Concurrency and Computation: Practice and Experience}, vol.~31, no.~16, p. e4964, 2019.

\bibitem{cooley1965algorithm}
J.~Cooley and J.~Tukey, ``An algorithm for the machine computation of the complex fourier series, in mathematics of computation.'' \emph{April}, 1965.

\bibitem{patel2021scalable}
S.~Patel, B.~Dave, S.~Kumbhani, M.~Desai, S.~Kumar, and B.~Chaudhury, ``Scalable parallel algorithm for fast computation of transitive closure of graphs on shared memory architectures,'' in \emph{2021 IEEE/ACM 6th International Workshop on Extreme Scale Programming Models and Middleware (ESPM2)}.\hskip 1em plus 0.5em minus 0.4em\relax IEEE, 2021, pp. 1--9.

\bibitem{moustafa2016datalography}
W.~E. Moustafa, V.~Papavasileiou, K.~Yocum, and A.~Deutsch, ``Datalography: Scaling datalog graph analytics on graph processing systems,'' in \emph{2016 IEEE International Conference on Big Data (Big Data)}.\hskip 1em plus 0.5em minus 0.4em\relax IEEE, 2016, pp. 56--65.

\bibitem{kumar:hipc:2019}
S.~Kumar and T.~Gilray, ``Distributed relational algebra at scale,'' in \emph{International Conference on High Performance Computing, Data, and Analytics (HiPC)}.\hskip 1em plus 0.5em minus 0.4em\relax IEEE, 2019.

\bibitem{gilray:cc:2021}
T.~Gilray and S.~Kumar, ``Compiling data-parallel datalog,'' in \emph{International Conference on Compiler Construction}.\hskip 1em plus 0.5em minus 0.4em\relax IEEE, 2021.

\bibitem{fan2021exploring}
K.~Fan, K.~Micinski, T.~Gilray, and S.~Kumar, ``Exploring mpi collective i/o and file-per-process i/o for checkpointing a logical inference task,'' in \emph{2021 IEEE International Parallel and Distributed Processing Symposium Workshops (IPDPSW)}.\hskip 1em plus 0.5em minus 0.4em\relax IEEE, 2021, pp. 965--972.

\bibitem{kumar:isc:2020}
S.~Kumar and T.~Gilray, ``Load-balancing parallel relational algebra,'' in \emph{ISC High Performance}.\hskip 1em plus 0.5em minus 0.4em\relax IEEE, 2020.

\bibitem{Davis:2011:UFS:2049662.2049663}
T.~A. Davis and Y.~Hu, ``The university of florida sparse matrix collection,'' \emph{ACM Trans. Math. Softw.}, vol.~38, no.~1, Dec. 2011.

\bibitem{faraj2005automatic}
A.~Faraj and X.~Yuan, ``Automatic generation and tuning of mpi collective communication routines,'' in \emph{Proceedings of the 19th annual international conference on Supercomputing}, 2005, pp. 393--402.

\bibitem{venkata2012exploring}
M.~G. Venkata, R.~L. Graham, J.~Ladd, and P.~Shamis, ``Exploring the all-to-all collective optimization space with connectx core-direct,'' in \emph{2012 41st International Conference on Parallel Processing}.\hskip 1em plus 0.5em minus 0.4em\relax IEEE, 2012.

\bibitem{traff2014implementing}
J.~L. Tr{\"a}ff, A.~Rougier, and S.~Hunold, ``Implementing a classic: Zero-copy all-to-all communication with mpi datatypes,'' in \emph{Proceedings of the 28th ACM international conference on Supercomputing}, 2014.

\bibitem{jocksch2020optimised}
A.~Jocksch, N.~Ohana, E.~Lanti, V.~Karakasis, and L.~Villard, ``Optimised allgatherv, reduce\_scatter and allreduce communication in message-passing systems,'' \emph{arXiv preprint arXiv:2006.13112}, 2020.

\bibitem{jackson2004planned}
A.~Jackson and S.~Booth, ``Planned alltoallv a cluster approach,'' 2004.

\bibitem{plummer2004lpar}
M.~Plummer and K.~Refson, ``An lpar-customized mpi alltoallv for the materials science code castep,'' \emph{Technical Report, EPCC (Edinburgh Parallel Computing Centre)}, 2004.

\bibitem{bienz2022locality}
A.~Bienz, S.~Gautam, and A.~Kharel, ``A locality-aware bruck allgather,'' in \emph{Proceedings of the 29th European MPI Users' Group Meeting}, 2022, pp. 18--26.

\end{thebibliography}
